\documentclass[12pt,preprint]{aastex}
\usepackage{emulateapj5}

\usepackage{graphicx}

\newcommand{\cmc}{\mbox{$\mbox{cm}^{-3}$}}
\newcommand{\kms}{\mbox{km$\,$s$^{-1}$}}
\newcommand{\cmg}{\mbox{$\mbox{cm}^{2} \, \mbox{g}^{-1}$}} 
\newcommand{\lsun}{\mbox{$L_\odot$}}
\newcommand{\msun}{\mbox{$M_\odot$}}
\newcommand{\jyb}{\mbox{$\mbox{Jy} \, \mbox{beam}^{-1}$}}
\newcommand{\htwo}{\mbox{$_{\mbox{\tiny H2}}$}}
\newcommand{\K}{\mbox{K}}
\newcommand{\hii}{H\mbox{\sc ~ii} }       %\ion{H}{2} not working
\newcommand{\hcop}{HCO$^+$}
\newcommand{\htcop}{H$^{13}$CO$^+$}
\newcommand{\hcops}{HCO$^+$ }
\newcommand{\htcops}{H$^{13}$CO$^+$ }
\newcommand{\lbol}{\mbox{$L_{\mbox{\tiny bol}}$}}
\newcommand{\lsmm}{\mbox{$L_{\mbox{\tiny $\lambda> 350~\micron$}}$}}
\newcommand{\speak}{\mbox{$S_{\mbox{\tiny 1.3~mm}}^{\mbox{\tiny ~peak}}$}}
\newcommand{\sint}{\mbox{$S_{\mbox{\tiny 1.3~mm}}^{\mbox{\tiny ~int}}$}}
\newcommand{\sff}{\mbox{$S_{\mbox{\tiny 1.3~mm}}^{\mbox{\tiny ~free-free}}$}}
\newcommand{\sintsharc}{\mbox{$S_{\mbox{\tiny 350~\micron}}^{\mbox{\tiny ~int}}$}}
\newcommand{\spindex}{\mbox{$\alpha_{\mbox{\tiny{350}}}^{\mbox{\tiny{1300}}}$}}
\newcommand{\tdust}{\mbox{$T_{\mbox{\tiny dust}}$}}
\newcommand{\kmm}{\mbox{$\kappa_{\mbox{\tiny 1.3~mm}}$}}
\newcommand{\Bmm}{\mbox{$B_{\mbox{\tiny 1.3~mm}}$(\tdust)}}
\newcommand{\msmm}{\mbox{$M_{\mbox{\tiny smm}}$}}
\newcommand{\mvir}{\mbox{$M_{\mbox{\tiny vir}}$}}
\newcommand{\vlsr}{\mbox{$v_{\mbox{\tiny LSR}}$}}
\newcommand{\tmb}{\mbox{$T_{\mbox{\tiny MB}}$}}
\newcommand{\eff}{\mbox{$\eta_{\mbox{\tiny MB}}$}}

\slugcomment{To appear in ApJ}

\shorttitle{Massive Star Formation in the W43 Complex}
\shortauthors{Motte et al.}

\begin{document}

\title{From Massive Protostars to a Giant H~{\sc II} Region: 
       Submillimeter Imaging of the Galactic Mini-starburst W43}

\author{F. Motte\altaffilmark{1}}
\affil{California Institute of Technology, Downs Laboratory of Physics,
       Mail Stop 320-47, 1200 E California Blvd, Pasadena, CA 91125, USA}
\email{motte@submm.caltech.edu}
\author{P. Schilke}
\affil{Max-Planck-Institut f\"ur Radioastronomie,
       Auf dem H\"ugel 69, 53121 Bonn, Germany}
\and
\author{D. C. Lis}
\affil{California Institute of Technology, Downs Laboratory of Physics,
       Mail Stop 320-47, 1200 E California Blvd, Pasadena, CA 91125, USA}

\altaffiltext{1}{Previous institute: Max-Planck-Institut f\"ur 
                 Radioastronomie, Auf dem H\"ugel 69, 53121 Bonn,
                 Germany}

\begin{abstract}
We have carried out a submillimeter continuum and spectroscopic study
of the W43 main complex, a massive star-forming region, which harbors
a giant \hii region.  The maps reveal a filamentary structure
containing $\sim 50$ fragments with masses of $40-4\,000~\msun$ and
typical diameters of 0.25~pc.  Their large sizes, large non-thermal
velocities ($\Delta v \sim 5~\kms$), and high densities ($n\htwo\sim
10^6~\cmc$) suggest that they are protoclusters and excellent sites to
form massive stars.  Follow-up observations are necessary, but we have
already identified three protoclusters to be very good candidates for
containing very young massive protostars.  The starburst cluster, that
excites the giant \hii region has a large impact on the molecular
complex.  However, it remains unclear if this first episode of star
formation is triggering the formation of new massive stars, through
ionization shocks crossing the closeby molecular clouds.  W43 is thus
an ideal laboratory to investigate massive star formation from the
protostellar phase to that of giant \hii regions.  Moreover, the very
active star-forming complex W43 may be considered a Galactic
mini-starburst region that could be used as a miniature model of
starburst galaxies.
\end{abstract}



%%%%%%%%%%%%%%%%%%%%%%%%%%%%%%%%%%%%%%%%%%%%%%%%%%%%%%%%%%%%%%%%%%%%
%% 1. Introduction
\section{Introduction}

High-mass (OB; $M_\star>8~\msun$) stars are believed to form in
clusters within molecular cloud complexes.  Due to their high
ultraviolet luminosity, massive young stellar objects (YSOs) first
heat, then ionize, and disrupt their surrounding molecular cloud.
Owing to their large distances to the Sun and the complex interplay
between massive YSOs and the neighboring interstellar medium, the
formation of high-mass stars is still poorly understood.  Many studies
have been devoted to embedded YSOs that have already developed an \hii
region and thus are easily detectable in far-infrared and centimeter
continuum surveys (see a review by \citealt{chur99}).  In contrast,
possible precursors of ultracompact \hii regions (UCH\mbox{\sc ~ii}s),
i.e. massive YSOs in their main building phase, have only been
discovered recently (\citealt{hunt00, bran01, srid02} and references
therein).  These massive protostars, also called high-mass
protostellar objects (HMPOs), are inconspicuous in the IRAS bands and
have no, or very little free-free emission.  They can be detected in
(sub)millimeter dust continuum and molecular high-density tracers and
frequently display H$_2$O/CH$_3$OH maser emission.  Therefore,
submillimeter continuum imaging of molecular cloud complexes is ideal
to probe clouds surrounding UCH\mbox{\sc ~ii}s and HMPOs
simultaneously.  Making a census of those deeply embedded phases is
the first necessary step to gain insight into the processes leading to
the formation of a massive star.

The W43 star-forming complex is located in the inner spiral arm of our
Galaxy, at 5.5~kpc from the Sun \citep{wils70}.  W43 is well-known for
its giant \hii region emitting $10^{51}$ Lyman continuum photons per
second and a far-infrared continuum luminosity of $\sim 3.5\times
10^6~\lsun$ \citep*{smit78, lest85}.  The main ionizing source of W43
was discovered in near-infrared images by \citet{lest85} and confirmed
by \citet*{blum99} as a cluster of Wolf-Rayet (WR) and OB main
sequence stars.  The inner 20~pc of W43 (at $l\sim 30.75\degr$, $b\sim
-0.06\degr$) contain this WR/OB cluster and a $10^6~\msun$ molecular
cloud called G30.8-0.0.  The W43 main complex was mapped in CO lines
and parts of it in higher density tracers such as H$_2$CO and \hcops
lines, and 1.3~mm continuum \citep*{bieg82, lisz95, moon95}.  Several
sources were identified by these authors but a complete and
comprehensive sample of UC\hii regions and massive protostars is still
lacking.  Identifying such a sample is essential for the present
paper, which aims at presenting a global scenario of (massive) star
formation in W43.  Therefore, we will carefully investigate the
distribution of dense molecular clouds in the W43 main complex, in
particular to identify sites of future or on-going massive star
formation.

A detailed study of W43 may help constrain the properties of distant
starburst galaxies.  Indeed, the large luminosity and ionizing flux of
the W43 giant \hii region are similar to those of NGC~3603 or M17
($10^5-10^7~\lsun$ and $10^{50}-10^{51}$~Lyc$\,$s$^{-1}$) , taken to
be representative of clusters and \hii regions in starburst galaxies
(e.g. \citealt{tapi01}).  The NGC~3603 stellar cluster is qualified as
``starburst'' because it consists of several tens of WR, O, and B-type
stars (e.g. \citealt{bran99}).  The W43 cluster of main sequence stars
is likely to be similar but more heavily reddened ($A_{\rm v}=30$~mag
versus 4~mag in NGC~3603).  Since the W43 stellar cluster is closely
associated with giant molecular clouds, we may be witnessing another
burst of star formation.  This would make the W43 main complex a
Galactic mini-starburst region, i.e. a miniature model of the stellar
and gas content of starburst regions in distant galaxies.

In the present paper, we report a submillimeter continuum and
spectroscopic study of the W43 main complex.  From the imaging and
deep spectroscopic measurements presented in Sect.~2, we make a
complete census of the compact and dense cloud fragments in W43
(Sect.~3).  In Sect.~4, we investigate the impact of the giant \hii
region on the molecular complex, the nature of these cloud fragments,
and the global characteristics of the mini-starburst W43.  Finally,
Sect.~5 summarizes our conclusions.

%%%%%%%%%%%%%%%%%%%%%%%%%%%%%%%%%%%%%%%%%%%%%%%%%%%%%%%%%%%%%%%%%%%%
%% 2. Observations
\section{Observations}

%% 2.1 Continuum
\subsection{Dust Continuum Observations}

We mapped the dust continuum emission of the W43 main complex at
1.3~mm and $350~\micron$ in February and April 1999.  We used the
37-channel MPIfR MAMBO bolometer array installed at the IRAM\footnote{
   IRAM is supported by INSU/CNRS (France), MPG (Germany) and IGN
   (Spain).}
30-meter telescope on Pico Veleta (Spain) to make the 1.3~mm map. We
performed the $350~\micron$ observations at the CSO\footnote{
   The CSO is operated by the California Institute of Technology under
   funding by the National Science Foundation, Contract AST-9615025.}
10.4-meter radiotelescope on Mauna Kea (Hawaii), equipped with the
24-pixel SHARC bolometer camera.  The passband of the MAMBO
(respectively SHARC) bolometer array has an equivalent width of
$\approx 70$~GHz (resp. 100~GHz) and is centered at $\nu_{\rm eff}
\approx 240$~GHz (resp. 860~GHz) \citep*{krey98, hunt96}.

Both 1.3~mm and $350~\micron$ images cover $\sim 12\arcmin \times
10\arcmin$ and were obtained by combining partially overlapping
on-the-fly maps.  In the dual-beam on-the-fly mapping mode, the
telescope is scanned continuously in azimuth along each row, while
wobbling.  For each channel, the raw data corresponding to a single
on-the-fly coverage consist of several rows taken at a series of
elevation offsets. For MAMBO (respectively SHARC) maps, we used a
scanning velocity of $4\arcsec\,$sec$^{-1}$ (resp.
$10\arcsec\,$sec$^{-1}$), and a sampling of $2\arcsec$ (resp.
$4\arcsec$) in azimuth and $4\arcsec$ in elevation.  The wobbler
frequency was set to 2~Hz (resp. 4~Hz) and the wobbler throw in
azimuth was $70\arcsec$ (resp. $135\arcsec$).  The typical azimuthal
size of individual maps was $\sim 9\arcmin$ with MAMBO versus $\sim
12\arcmin$ with SHARC.  The MAMBO maps were reduced with the IRAM
software for bolometer-array data (NIC; cf. \citealt*{brog95}) while
the SHARC data were processed with the standard CSO programs CAMERA
and REGRID. The IRAM and CSO programs both use the EKH restoration
algorithm \citep*{EKH}.

The size of the main beam was measured to be $HPBW \sim 11\arcsec$ for
MAMBO maps and $\sim 11.5\arcsec$ for SHARC maps using Uranus and
Mars.  The absolute pointing of each telescope was found to be
accurate to within $\sim 5\arcsec$.  The data were taken with moderate
weather conditions. The zenith atmospheric optical depth at 240~GHz
(respectively 225~GHz) varied between $\sim 0.2$ and $\sim 0.4$ for
MAMBO observations and was roughly $\sim 0.06$ for SHARC observations.
Uranus and Mars were used for flux calibration and the overall,
absolute calibration uncertainty is estimated to be $\sim 20\%$ for
both MAMBO and SHARC maps.

%% 2.2 Lines
\subsection{Molecular Line Observations}

In June 2000 and 2001, we mapped the \hcop(3-2) rotational line
emission of the W43 main complex and conducted deep integrations at
some of the most prominent dust maxima in the (3-2) transitions of
\hcops and its \htcops isotopomer.  We used the facility 230~GHz
receiver of the CSO and its 50~MHz bandwidth acousto-optical
spectrometer with a $\sim 0.16~\kms$ spectral resolution.  The CSO
{\it FWHM} angular resolution for frequencies of 267.5576~GHz and
260.2555~GHz is $28\arcsec-29\arcsec$.

We made two overlapping on-the-fly maps that cover a $8\arcmin \times
10\arcmin$ area.  Each of these maps consists of several rows
continuously scanned in right ascension and taken at a series of
declination offsets.  The scanning velocity was $7\arcsec\,$sec$^{-1}$
and the sampling $10\arcsec$ in both directions.  The off position was
taken at each declination offset, $20\arcmin$ away in right ascension.
Besides this map, we performed deep measurements in position switching
mode, using off positions $\pm 20\arcmin$ in azimuth.  Pointing was
checked on planets when available or W-AQL otherwise and was found to
be stable to within $5\arcsec-10\arcsec$.

The data were reduced with the IRAM software for spectral lines
(CLASS), applying low order baselines and averaging individual spectra
with rms weighting.  Our line observations were corrected for
atmospheric attenuation, ohmic losses, rearward spillover and
scattering, using the standard chopping wheel technique.  The 225~GHz
zenith optical depth was $\sim 0.075$ during the maps and varying
between 0.08 and 0.18 during the position switch measurements.  To
convert the resulting antenna temperatures ($T_{\mbox{\tiny
A}}^\star$) to main beam brightness temperatures (\tmb), we have
divided the data by the main beam efficiency determined from
observations of Mars, $\eff=0.65$.  The calibration accuracy is $\sim
25\%$.

%%%%%%%%%%%%%%%%%%%%%%%%%%%%%%%%%%%%%%%%%%%%%%%%%%%%%%%%%%%%%%%%%%%%
%% 3. Results and analysis
\section{Results and Analysis}

Our 1.3~mm and $350~\micron$ continuum maps are presented in
Figs.~\ref{f:dust}a-b.  They cover a 17~pc~$\times$~20~pc area,
roughly centered on the stellar WR/OB cluster.  The \hcop(3-2) map
covers a slightly smaller area (14~pc~$\times$~16~pc) and is shown in
Figs.~\ref{f:hcop}a-f, where the integrated intensity is displayed
along with five velocity channels ranging from $86~\kms$ to
$102~\kms$.  Figs.~\ref{f:dust}a-b and \ref{f:hcop}a all show that the
dense molecular gas in the W43 main complex follows a global
``Z''-shaped filamentary structure.  Some more compact fragments are
observed within this filament.  We make a complete census of these
fragments in Sect.~\ref{s:census} and estimate their temperature,
mass, and other characteristics in Sect.~\ref{s:mass}.

%% 3.1 Census
\subsection{Census of Compact Cloud Fragments in W43}\label{s:census}

%% 3.1.1 Structure
\subsubsection{Cloud structure}\label{s:structure}

Our immediate goal is to investigate the submillimeter continuum maps
of W43 to identify cloud fragments which could form massive stars.
The maps shown in Figs.~\ref{f:dust}a-b are in very good agreement
with each other (except for the north-western corner, see below) and
mainly trace the optically thin thermal emission of dust in molecular
clouds.  Examination of Figs.~\ref{f:dust}a-b reveals a wealth of
cloud structures with size scales ranging from $11\arcsec$ ($\sim
0.3$~pc) to more than $500\arcsec$ ($\sim 13$~pc).  The dynamical
range within these maps is also very high: $450\sigma$ in the MAMBO
map and $75\sigma$ in the SHARC map.  To assess the complexity of the
submillimeter emission of W43, we use the Gaussclumps program
\citep{kram98}, which we have adapted to continuum observations.  The
emission above $5\sigma=60~\jyb$ in Fig.~\ref{f:dust}a can fairly well
be described by a sum of $\sim 300$ cloud fragments with 2-D Gaussian
shapes.  Since most of such fragments consist of moderate-density gas
($n\htwo< 10^4 ~\cmc$, see below), they are unlikely sites for massive
star formation.

We thus need to separate compact and dense cloud fragments from their
surrounding, lower density clouds.  We use the method developed by
\cite{ma01a} to estimate the outer diameters of the four main
submillimeter fragments (W43-MM1 to MM4 labelled in
Fig.~\ref{f:dust}a).  In both MAMBO and SHARC maps, we measure outer
diameters of $\sim 40\arcsec$ corresponding to $\sim 1$~pc.  We do not
expect the weaker compact fragments to have the exact same radial
extent but, for simplicity, we use a $40\arcsec$ lengthscale to
extract all dense fragments.  We stress that such a cloud
decomposition is not unique as these $\sim 1$~pc cloud fragments will
probably fragment further to form more than one single star.  It
permits one to select structures with peak density larger than $\sim
10^4~\cmc$, when a $5\sigma = 60$~mJy/11\arcsec-beam 1.3~mm flux, a
20~K dust temperature and a $\kmm=0.01~\cmg$ dust mass opacity are
assumed (calculated from e.g. Eq.~($1'$) of
\citealt*{mott98}).

We measure the submillimeter flux of each compact fragment in two
steps.  We first filter out all spatial scales larger than $40\arcsec$
from the MAMBO map and identify 51 fragments above the $5\sigma$
level.  We then derive the 1.3~mm Gaussian characteristics of each
fragment by applying an improved version of the technique we developed
for our previous studies \citep{mott98, mott01}.  It uses a
multiresolution analysis to refine the extraction of the compact
fragment from its local background emission (see \citealt*{star98,
ma01}) and the Gaussclumps program to fit its Gaussian parameters
\citep{kram98}.  The fitted fragments are indicated as black ellipses
in Fig.~\ref{f:dust}a and account for $\sim 15\%$ of the total 1.3~mm
flux measured.  We perform the same analysis on the SHARC map, where
23 of the 51 1.3~mm fragments are recovered above $3\sigma=9~\jyb$ and
within $10\arcsec$ of the 1.3~mm positions.

These compact submillimeter fragments have basic characteristics
listed in Table~\ref{t:clumps}.  Col.~1 gives their adopted names
(where a small number corresponds to a high level of detection during
the first step of our analysis) and Cols.~2-3 their coordinates.  The
Gaussian peak fluxes, {\it FWHM} sizes, and integrated fluxes measured
at 1.3~mm are listed in Cols.~4-6, along with the integrated fluxes of
their $350~\micron$ counterparts in Col.~7.  The spectral indices
$\spindex$ (where $S_\nu \propto \nu^\alpha$) we measure from the
integrated fluxes $\sint$ and $\sintsharc$ are given in Col.~8.  The
error bar on $\spindex$ varies from $\pm 0.1$ up to $\pm 0.4$ for the
weaker fragments.

The next step of the analysis is to ensure that the submillimeter
fragments selected above are indeed dense cloud fragments.  We thus
need to check the contamination of submillimeter fluxes by lines and
free-free emission.  Contamination by CO line emission is negligible
for those fragments which are dense and compact.  In contrast, the
``hot core'' line emission displayed by several fragments (Motte et
al. in prep.) contributes to the submillimeter continuum fluxes we
measure.  Such ``hot cores'' are pointlike (in a 0.3~pc beam) and
embedded in dense fragments, which have a $\sint/\speak$ ratio of
$\sim 2$ (cf. Table~\ref{t:clumps}).  Their emission would thus very
unlikely dominate the submillimeter continuum.  We consider that
$25\%$ is an upper limit for the line contamination of the W43
submillimeter fragments.

Without a doubt, the main source of contamination of 1.3~mm fluxes is
the free-free emission from \hii regions.  To estimate such a
contamination, we compare our 1.3~mm and $350~\micron$ images with
centimeter free-free data from the literature.  Strikingly, the maps
of Figs.~\ref{f:dust}a-b differ in their north-western corner where
the 1.3~mm arc-like structure disapears at $350~\micron$.  For a
further comparison we compute the spectral index map between
$350~\micron$ and 1.3~mm (see Fig.~\ref{f:index}).  Most of the map
shows an index of $\spindex\sim 4 \pm 0.5$, suggestive of thermal
emission of dust at $\ge 30~\K$.  In contrast, at the location of the
giant \hii region \citep*{lest85, lisz93, bal01}, the map has a lower
index: $\spindex \sim 2.5-3.5$.  Since there is a good correspondance
between these low $\spindex$ values and the large-scale ionized gas
(cf. Fig.~\ref{f:index}), the 1.3~mm fluxes measured in that region
are likely contaminated by free-free emission.  Most of the free-free
emission in W43 arises from its giant \hii region and is probably
optically thin at centimeter wavelengths.  Any ultra-compact \hii
regions in W43 should have a free-free emission at least partly
optically thick.  The spectral indices computed for the two most
compact centimeter sources are however indicating that their free-free
emission is mostly optically thin when $\lambda < 3$~cm (see
Sect.~\ref{s:crosscorr}).  We thus hereafter assume that all the
3.5~cm emission mapped in W43 at $\sim 10\arcsec$ angular resolution
is optically thin.

We estimate the contribution of free-free emission to the 1.3~mm
fluxes measured in Table~\ref{t:clumps} by using the 3.5~cm map of
\citet{bal01}, which has a similar angular resolution to our continuum
data ($9\arcsec$ vs. $11\arcsec$).  The 21~cm map of \cite{lisz93} is
used when the fragments are lying outside the 3.5~cm primary beam.  We
filter out all spatial scales larger than $40\arcsec$ and integrate
the centimeter emission at the position of each 1.3~mm fragment.  Then
we assume a $S_\nu \propto \nu^{-0.1}$ spectra typical for optically
thin free-free emission to estimate the flux of the ionized gas at
1.3~mm (cf. Col.~9 in Table~\ref{t:clumps}).  Within the giant \hii
region powered by the WR/OB cluster, the free-free contribution to the
twenty 1.3~mm compact fragments (including W43-MM6, MM8, MM14, MM11,
MM4, MM20, MM15 and MM25 shown in Fig.~\ref{f:dust}a) goes up to
$70\%$ with an average value of $16\%$.  Outside this region, the
free-free contribution is lower than $10\%$, with the notable
exception of W43-MM13, which coincides with a centimeter source
(cf. Fig.~\ref{f:index}).  Altogether the level of free-free
contamination is low for the compact continuum fragments identified in
Fig.~\ref{f:dust}a.

In the following, we subtract the free-free emission estimated above
from the 1.3~mm fluxes of Table~\ref{t:clumps}.  Three of the 51
fragments (namely W43-MM32, MM46, MM50) are falling below the
$5\sigma$ limit we set for their initial detection and are removed
from the sample.  We thus end up with 48 1.3~mm fragments, 22 of which
are also detected at $350~\micron$.  We are confident that the 48
compact fragments identified in Table~\ref{t:mass} represent genuine
dust continuum sources.

%% 3.1.2 Kinematics
\subsubsection{Cloud kinematics}\label{s:kinematics}

We make here a first order kinematics analysis of the dense cloud
fragments observed in the W43 main molecular complex.  The \hcop(3-2)
emission in Fig.~\ref{f:hcop}a displays the same filamentary structure
as the dust continuum maps in Figs.~\ref{f:dust}a-b.  We selected
several compact dust fragments at various locations in the W43 main
complex and made deep integrations in \hcop(3-2) and \htcop(3-2).  In
Table~\ref{t:lines}, for each submillimeter fragment (Col.~1) and
optically thin molecular transition (Col.~2), we separate the line
into one to four velocity components (see Col.~3).  We choose this
decomposition to better describe the lines which are broad and
structured but none of those components are perfectly fitted by a
Gaussian.  We give in Cols.~4-7 of Table~\ref{t:lines} the Gaussian
parameters of the fitted components, i.e. local velocity at rest, peak
line temperature, integrated intensity and line width.
Figs.~\ref{f:line}a-e display the spectra taken at some of these
locations. \hcops emission has been successfully detected for each
selected submillimeter fragment.  Even 1.3~mm locations W43-MM15\&20
and MM25, where heating from the WR/OB cluster and contamination by
free-free emission is highly probable, contain high-density gas.  As
suggested by a careful examination of Fig.~\ref{f:hcop}a , the
\hcop(3-2) line emission of W43-MM1, MM2, MM3, MM9, and MM10 is
optically thick (cf. Figs.~\ref{f:line}a and
\ref{f:line}e).  The \hcop(3-2) emission of the other submillimeter
fragments is probably optically thin, but their lines often display a
complex shape: from multiple velocity components (W43-MM15\&20, MM4,
MM7, cf. Figs.~\ref{f:line}c-d) to a single component with a broad
non-Gaussian line shape (W43-MM6\&8, MM25, cf. Fig.~\ref{f:line}b).
We stress that the more complex is the line, the closer in projection
to the WR/OB cluster is the fragment.  Despite those limitations, we
will use the values of $\vlsr$ and $\Delta v$ given in
Table~\ref{t:lines} to further analyze the molecular cloud kinematics.

As illustrated in Figs.~\ref{f:hcop}b-f and Figs.~\ref{f:line}a-d,
there is a velocity gradient from the southern part of the filament
($\vlsr\sim 85~\kms$) to its northern part ($\vlsr\sim 105~\kms$).
\citet{lisz95} observed similar gradients in \hcop(1-0) and
$^{13}$CO(1-0) and proposed that the southern gas is in front of W43
and the northern gas behind (see also \citealt{bieg82}). This velocity
gradient roughly develops along the Galactic plane and encompasses the
recombination line velocity found for the ionized gas in front of the
WR/OB cluster ($92.4~\kms$ according to \citealt{bal01}).  It thus
suggests that the starburst cluster is intimately associated with the
W43 main molecular complex.

%% 3.2 Masses
\subsection{Characteristics of the Compact Cloud Fragments}\label{s:mass} 

We derive the basic properties of the compact ($< 1$~pc) cloud
fragments of W43 using their dust continuum and \hcops line parameters
given in Tables~\ref{t:clumps}-\ref{t:lines}.  Table~\ref{t:mass}
lists the fragment name (Col.~1), the adopted dust temperature
(Col.~2), the estimated submillimeter mass (Col.~3), gas density
(Col.~4), and virial mass (Col.~5).  In the following, we describe the
methods and assumptions used to estimate those characteristics and we
search for stellar activity signatures coincident with the dense cloud
fragments.

%% 3.2.1 Temperature
\subsubsection{Dust temperature}\label{s:temp} 

The temperature is unlikely to be homogeneous in the W43 molecular
complex because it is known to form massive stars.  As a matter of
fact, \cite{lest85} measured dust temperatures ranging from 40~K to
90~K, using $50~\micron$ and $100~\micron$ images made at the Kuiper
Airborne Observatory (KAO).  We expect the temperature of W43 compact
fragments to be lower as they are largely embedded within clouds.  The
first confirmation comes from the spectral indices of the 22 fragments
detected at both 1.3~mm and $350~\micron$.  When they are corrected
for free-free contamination, these indices have a median value of
$\spindex\simeq 3.3$ (see Cols.~8-9 of Table~\ref{t:clumps}).  For a
pure thermal and optically thin dust emission with a standard dust
opacity index of $\beta= 2$ (where $\kappa_\nu \propto \nu^\beta$),
this corresponds to the averaged dust temperature of $\sim 20~\K$.

In an effort to better constrain the average temperature of the W43
compact fragments we compare their spectral energy distribution (SED)
to graybody models (cf. Figs.~\ref{f:sed}a-b).  Their SED is built
from the 1.3~mm and $350~\micron$ integrated fluxes measured in this
paper (cf. Table~\ref{t:clumps}) and the $100~\micron$ and
$50~\micron$ fluxes read from Fig.~1 of \cite{lest85} for pointlike
sources (in a $50\arcsec$ beam).  The fluxes of MSX sources located
close to the submillimeter fragments or corresponding upper limits are
displayed but not used (see Sect.~\ref{s:crosscorr}).  Our SED
analysis shows that graybody models with dust opacity index of
$\beta=2$ might be more appropriate than those with $\beta=1.5$ (where
$\kappa_\nu \propto \nu^\beta$, cf. Fig.~\ref{f:sed}a).  This result
agrees with the multiwavelength submillimeter studies of other massive
YSOs (e.g. \citealt{hobs93}).  We therefore use graybody models with
$\beta=2$, while knowing that temperatures increase by only a few
degrees when $\beta=1.5$ (e.g.  Fig.~\ref{f:sed}a).  Since mid- to
far-infrared fluxes are often missing and upper limits remain high
(see also Sect.~\ref{s:giant}), we only successfully constrain the
dust temperature of four dense fragments: W43-MM1 ($\tdust \sim
19~\K$), MM2 ($\sim 23~\K$), MM3 ($\la 20~\K$), and MM7 ($\la 22~\K$).
We also derive the bolometric luminosity and the submillimeter to
bolometric luminosity ratio of W43-MM1 ($\lbol
\sim 2.3\times 10^4~\lsun$, $\lsmm/\lbol \sim 3\%$), MM2 ($\sim
2.4\times 10^4~\lsun$, $\sim 1.5\%$), MM3 ($\la 1.2\times 10^4~\lsun$,
$\ga 2.5\%$), and MM7 ($\la 1.0\times 10^4~\lsun$, $\ga 1.5\%$).
Despite its incompleteness, the present analysis suggests that most
compact cloud fragments in W43 are rather cold ($\tdust \sim 20~\K$)
and have high bolometric luminosities ($\lbol \sim 10^4~\lsun$).
Their masses are estimated below.

%% 3.2.2 Submm Mass
\subsubsection{Mass estimated from the submillimeter continuum}\label{s:smmmass} 

We believe that the submillimeter continuum emission of cloud
fragments identified in Sect.~\ref{s:structure} is mainly thermal dust
emission, which is largely optically thin.  For any given dust
properties and gas-to-dust ratio, the 1.3~mm and $350~\micron$ fluxes
are thus directly related to the total (gas~$+$~dust) mass of the
fragments.  For present mass estimates, we use the 1.3~mm integrated
fluxes ($\sint$, see Table~\ref{t:clumps}) because they provide more
homogeneous measures.  We correct those fluxes for any residual
free-free contamination (cf. Col.~9 in Table~\ref{t:clumps}) and
derive the ``submillimeter mass'' (\msmm, given in
Table~\ref{t:mass}), as follows:
\begin{eqnarray}
\msmm & =    & \frac{(\sint)^{\mbox{\tiny corr}}\; d^{2}}{\kmm \; \Bmm} \nonumber\\
      &\simeq& 5.5\ \msun \times 
              \left(\frac {(\sint)^{\mbox{\tiny corr}}}{\mbox{0.01~Jy}}\right) 
              \left(\frac {d}{\mbox{5.5~kpc}} \right)^2 \nonumber\\
      &      & \times \left(\frac {\kmm}{0.01\,\cmg}\right)^{-1}
                     \left(\frac {\tdust}{20~\K}\right)^{-1},
\label{eq:mass}
\end{eqnarray}
where $\kmm$ is the dust opacity per unit mass column density at
1.3~mm, and $\Bmm$ is the Planck function for a dust temperature
\tdust.

The dust mass opacity (including dust properties and gas-to-dust mass
ratio) is likely to vary with density, temperature, and the
evolutionary state of the emitting medium \citep*{henn95}.  Models of
dust in low-mass protostellar cores (e.g. \citealt{osse94}) suggest
that a value of $\kmm=0.01~\cmg$ is well suited for cool ($10-30~\K$)
and high-density ($n\htwo\ga 10^5~\cmc$) cloud fragments.  This value
agrees with the recent cross-comparisons of dust emission surrounding
an UC{\hii} region with its CO ice absorption and gas emission
\citep{vdt02}.  We thus choose a dust opacity per unit (gas~$+$~dust)
mass column density of $\kmm=0.01~\cmg$.  We use this value for all
the dense fragments because the average dust properties of these $\sim
0.25$~pc structures should not change drastically when they contain a
massive protostar or an UCH\mbox{\sc ~ii}.  We estimate that the
absolute value taken for the dust mass opacity is uncertain by at
least a factor of 2.

The temperature to be used in Eq.~(\ref{eq:mass}) is the mass-weighted
dust temperature of the cloud fragments, its value can be determined
from the gray-body fitting of their SED.  In Sect.~\ref{s:temp}, we
recommend $20~\K$ for most of the dense cloud fragments.  We thus
assume $\tdust=20~\K$ for the fragments deeply embedded into the
``Z''-shaped filament and $\tdust=30~\K$ for those that may be less
shielded against the radiation of the WR/OB cluster (see Col.~2 of
Table~\ref{t:mass}).

%% 3.2.3 virial mass
\subsubsection{Virial mass}\label{s:virmass}

We estimate the virial mass of the W43 compact cloud fragments
assuming that they are spherical and have a $\rho\propto r^{-2}$
density distribution: $\mvir=3\, R \sigma^2/G$ (e.g. \citealt{bert92},
cf. Table~\ref{t:mass}).  We use the line width ($\Delta v$) listed in
Table~\ref{t:lines} and the diameter measured at 1.3~mm ({\it FWHM} in
Table~\ref{t:clumps}).  We subtract the thermal component of the line
width using the temperature adopted in Table~\ref{t:mass}.  As the
thermal component of the velocity dispersion is rather small
($a_{\mbox{\tiny s}}\sim 0.3~\kms$ when $\Delta v\sim 5~\kms$), the
error introduced by assuming that the kinetic temperature equals the
dust temperature is negligible.  Whenever the optically thin line
consists of several velocity components, we use their median line
width and indicate the number of components in Col.~5 of
Table~\ref{t:mass}.

The targetted positions for \hcops measurements generally correspond
to a single submillimeter fragment embedded in a lower-density
filament (cf. Sect.~\ref{s:structure}).  Therefore the line width and
then the virial mass derived from these measurements are upper limits
for the width and mass of the submillimeter fragments alone.
Follow-up maps with higher density tracers are needed to select the
gas only associated with the submillimeter fragment and estimate a
more accurate virial mass.

In a few cases, \hcops lines are optically thin but reveal several
velocity components.  The comparison with the line spectra of the
\htcops isotopomer can help isolate the velocity component most
probably associated with the compact and dense dust fragment.  This is
the case of W43-MM4 and MM7 where the low velocity components (at
$\vlsr\sim 91.8~\kms$ and $\sim 90.6~\kms$) have larger \htcop/\hcops
intensity ratios and should thus consist of higher density gas
(cf. Table~\ref{t:lines} and Fig.~\ref{f:line}d).  When we cannot
state that a submillimeter fragment corresponds to a single velocity
component, we simply assume that it consists of several cloud
fragments with the same mass.

According to \cite{pb93}, a fragment remains gravitationally bound as
long as $M/\mvir>0.5$.  Since the submillimeter masses are uncertain
by a factor of 2 and the virial masses are upper limits, we estimate
that the $\msmm/\mvir$ ratio must be at least larger than 0.2 to
suggest gravitational boundedness.

%% 3.2.4 Crosscorelation
\subsubsection{Coincidence with signposts of stellar activity}\label{s:crosscorr}

Various published catalogues can be used to search for stellar
activity within the dense cloud fragments listed in
Table~\ref{t:mass}.  In addition to the IRAS survey, the Midcourse
Space Experiment (MSX\footnote{See
http://www.ipac.caltech.edu/ipac/msx/msx.html.}) provides pointlike
sources observed at 8, 12, 15, and $21~\micron$ with a $20\arcsec$
resolution \citep{egan01}.  MSX G30.7196-0.0854 and G30.6882-0.0726
(or IRAS 18450-0205) are the only two infrared sources relevant to our
study because they coincide with high-density clouds and have
$<40\arcsec$ sizes (see Fig.~\ref{f:msx} and Sect.~\ref{s:giant}).
MSX G30.7196-0.0854 may be associated with W43-MM3, but its nominal
position is further away than what the maximum pointing uncertainties
allow.  As for MSX G30.6882-0.0726, it coincides with W43-MM13 well
within the $9\arcsec$ maximum positional errors.  The $60~\micron$ and
$100~\micron$ fluxes of the associated IRAS source are confused by the
far-infrared emission of the W43 giant \hii region.  With
HIRES\footnote{http://www.ipac.caltech.edu/ipac/iras/hires\_over.html}
(High Resolution Processing using the Maximum Correlation Method)
images, we successfully measure a $60~\micron$ flux of $1\,000 \pm
500$~Jy within a $\sim 60\arcsec$ beam.

Besides infrared sources, some small-diameter ($<40\arcsec$) radio
sources are identified at 21, 6 and 3.5~cm by \citet{zoot90},
\citet{garw88}, \citet{beck94}, and \citet{bal01}.  They roughly
coincide with W43-MM3, MM4, MM13, MM14 and MM20 (see
Fig.~\ref{f:index}).  W43-MM3 and MM4 are clearly the most compact of
these sources with diameters of $\sim 3\arcsec-5\arcsec$.  Their
spectral indices measured from 21 to 6~cm and then from 6 to 3.5~cm
are 0.5 and $\la 0.1$ for W43-MM3 and $\sim 0.6$ and $\la 0.5$ for
W43-MM4.  Their free-free emission is thus partially optically thick
from 21 to 3.5~cm and becomes mostly optically thin at wavelengths
shorter than 3.5~cm.

Furthermore, complete surveys of methanol, water and hydroxyl maser
emission have been performed in W43 \citep{wals98, vald01, braz83}.
The CH$_3$OH survey made at 6.7~GHz with ATCA has position
uncertainties smaller than $2\arcsec$ \citep{wals98} and the H$_2$O
maser positions observed at 22~GHz with the Medicina telescope are
only secure to $25\arcsec$ \citep{vald01}.  The OH type-I masers
detected at 1.6-1.7~GHz and compiled by \cite{braz83} have a
positional accuracy of $5\arcsec$ or $30\arcsec$ (south-west of
W43-MM2).  In Fig.~\ref{f:dust}b, W43-MM1 and MM11 are coincident with
the CH$_3$OH, H$_2$O and OH maser sources called Main3 and Main1,
respectively.  W43-MM2 is associated with CH$_3$OH and OH masers while
W43-MM6 and MM7 coincide with two H$_2$O maser sources, the second
being called Main2.  The position and systemic velocity of the
methanol masers are well confined within the radial and velocity
extent of the compact fragments ($<10\arcsec$ and $<3~\kms$ from their
peaks, \citealt{wals98}; see also \citealt{casw95}).  The water maser
coincident with W43-MM1 peaks $0.7~\kms$ away from its systemic
velocity.  That associated with W43-MM11 has one out of two velocities
recorded within $2~\kms$ of the molecular emission peak
(cf. \citealt{vald01} and references therein).  These infrared,
centimeter and maser sources are stellar activity signposts which will
help investigate the stellar content of cloud fragments listed in
Table~\ref{t:mass}.

%%%%%%%%%%%%%%%%%%%%%%%%%%%%%%%%%%%%%%%%%%%%%%%%%%%%%%%%%%%%%%%%%%%%
%% 4. Discussion
\section{Discussion}

We have made a complete census of cloud fragments with density greater
than $10^4~\cmc$ and size smaller than 1~pc in W43.  Before we discuss
their characteristics and evolutionary state (in
Sect.~\ref{s:protoclusters}), we investigate the likely effects of the
WR/OB cluster on the W43 molecular cloud complex (in
Sect.~\ref{s:WRcluster}).

%% 4.1 The WR cluster interaction
\subsection{The Starburst Cluster Interacting with W43 Molecular Clouds}\label{s:WRcluster}

According to the near-infrared spectroscopic study of \cite{blum99},
W43 harbors a cluster of main sequence stars that contains at least
one Wolf-Rayet star and two O-type (super)giants.  This young cluster
is substantially more luminous ($\sim 3.5\times 10^6~\lsun$) than a
typical OB association and qualifies as a starburst cluster.  We first
describe the characteristics of the giant
\hii region excited by this starburst cluster and then discuss its
impact on the W43 molecular clouds.

%% 4.1.1 Giant HII
\subsubsection{The giant \hii region}\label{s:giant}

Previous centimeter continuum studies showed that the far-ultraviolet
photons of the WR and OB stars in the W43 cluster created a large
bubble of free-free emission also called giant \hii region (see
\citealt{bal01} and references therein).  The white dashed ellipse
plotted in Fig.~\ref{f:index} outlines this bubble of ionized gas,
which has a full size of $\sim 4$~pc~$\times\: 6$~pc.  The
north-eastern and south-western extensions of the W43 free-free
emission may also be associated with gas ionized by the WR/OB cluster.
If this idea is confirmed, the \hii region powered by the W43
starburst cluster would cover up to 13~pc.

Fig.~\ref{f:msx} displays the infrared emission mapped with MSX at
$21~\micron$.  It nicely coincides with the free-free emission of the
giant \hii region (compare with Fig.~\ref{f:index}).  Since very small
grains \citep*{dese90} survive inside \hii regions, the 8 to 21~$\mu$m
MSX maps may trace their strong near- to mid-infrared continuum.
Indeed, this hot continuum has already been detected in a region with
similarly intense radiation field (namely M17), using ISOCAM/CVF
spectra at $5-16~\micron$ and MSX data \citep{cesa96, niel01}.  The
$21~\micron$ emission in Fig.~\ref{f:msx} peaks at the location of the
stellar cluster but also extends along the southern and northern
clouds containing W43-MM4, MM11, MM14 and W43-MM19, MM6, MM8, MM32.
Those ridges of mid-infrared emission may outline the ionization
fronts also observed at centimeter and far-infrared wavelengths
(cf. \citealt{lest85}).  The relative positions of the $21~\micron$
ridges to the 1.3~mm dense clouds perfectly agree with the 3D model of
W43 (cf. Fig.~\ref{f:msx}).  The southern $21~\micron$ emission is
displaced to the north of the clouds associated with W43-MM4, MM11 and
MM14, suggesting that most of it is obscured by these clouds where we
measure up to $A_{\rm v} \sim 300$~mag. In contrast, the northern
$21~\micron$ ridge is consistent with the illumination of the clouds
surface from the front and south.  The $10^5~\lsun$ luminosity and
$\sim 1$~pc extent of these ridges also agree with the radiation of
the cluster being intercepted within reasonable opening angles of
$1/20\times 4\pi$~sr.  It is thus very likely that the starburst
cluster, located in between these clouds (see
e.g. Fig.~\ref{f:vel_dist}a), is responsible for the large
mid-infrared emission mapped in e.g. Fig.~\ref{f:msx}.  Therefore, the
two IRAS sources 18450-0200 and 18449-0158 (from the Point Source
Catalogue) could arise at the interface between the ionized gas and
the molecular cloud rather than from any embedded stellar population
(see also \citealt{math90}).  This assumption holds when comparing the
IRAS fluxes with those measured by MSX and the KAO which both have
better angular resolution ($20\arcsec$ and $50\arcsec$ vs. $30\arcsec$
and $60\arcsec-120\arcsec$).  The 12 and $25~\micron$ fluxes of IRAS
18450-0200 and IRAS 18449-0158 are indeed twice stronger than those of
the compact sources MSX G30.7585-0.0495 and MSX G30.7815-0.0219
\citep{egan01}.  Similarly, the 60 and $100~\micron$ IRAS fluxes are
ten times those measured with the KAO (Figs.~1a-b in
\citealt{lest85}).  A near-infrared study will uniquely determine if
these two IRAS sources are indeed associated with ionization fronts
excited by the WR/OB cluster (Petr-Gotzens et al. in prep.).

The existing data suggest that the centimeter and near- to
far-infrared characteristics of W43 are completely dominated by the
ionization and heating from the WR/OB cluster.  We hereafter
investigate if the kinematics of the molecular clouds observed in
Figs.~\ref{f:dust}-\ref{f:hcop} is itself shaped by this starburst
cluster.

%% 4.1.2 Impact
\subsubsection{Impact of the starburst cluster on the clouds}\label{s:impact}

To distinguish between the influence of Galactic motions and that of
the starburst cluster, we need to carefully study the systemic
velocity throughout the W43 cloud complex.  Fig.~\ref{f:vel_dist}a
displays the $\vlsr$ measured for the compact cloud fragments of
Table~\ref{t:lines} as a function of the Galactic longitude offsets
relative to that of the starburst cluster.  It shows a clear velocity
gradient along the Galactic plane, which is consistent with that
mentioned in Sect.~\ref{s:kinematics}.  The dashed line in
Fig.~\ref{f:vel_dist}a represents a linear fit, which gives a gradient
of $\sim 0.4~\kms\,$pc$^{-1}$.  Except in the improbable case where
the W43 molecular cloud is along the line of sight ($i\la 3\degr$),
this velocity gradient is too large to be explained by a simple model
of spiral Galactic motions (e.g. \citealt{bran93}).  In contrast such
a gradient agrees with W43 being in the molecular ring, at a location
where the velocity field is known to be complex \citep{lisz93,
lisz95}.

In Fig.~\ref{f:vel_dist}b, the $\vlsr$ is corrected for the Galactic
gradient measured in Fig.~\ref{f:vel_dist}a and plotted as a function
of the projected distance to the WR/OB cluster.  This distance is set
to a negative value for cloud fragments situated south of the stellar
cluster.  The dispersion of the cloud fragments and the velocity
dispersion of individual fragments are observed to increase in the
inner 4~pc from the WR/OB cluster, outlined by a dashed ellipse in
Fig.~\ref{f:vel_dist}b.  Furthermore, the dispersion from the median
velocity (zero level indicated as a gray line) is larger for lower
density unbounded fragments.  Indeed, the four velocity components
detected at location W43-MM15\&20 have density lower than $10^5~\cmc$
and are far from virial equilibrium ($\msmm/\mvir \sim 0.04$).
Besides, the high velocity components of W43-MM4 and W43-MM7 probably
correspond to the lowest density clouds at the location of these dust
fragments (cf. Sect.~\ref{s:virmass}).  Therefore, we find that the
dispersion of the fragments systemic velocities is decreasing with the
distance to the WR/OB cluster and with the density of the clouds.  It
can be interpreted as evidence that low- to moderate-density clouds
($<10^5~\cmc$, $<50~\msun$) situated within a $3-4$~pc radius from the
starburst cluster have been blown away.  If we assume a typical age of
$10^6$~yr for the WR and OB stars, the $\sim 30~\kms$ dispersion
recorded on the line of sight of the cluster and supposed to
correspond to the 7~pc diameter measured in Fig.~\ref{f:vel_dist}b
suggests a sweeping motion of $\sim 3.5~\kms$.  A similar study around
the M17 giant \hii region suggests that a shell of cloud is radially
expanding with $\sim 10~\kms$ \citep{rain87}.  Such velocities can be
reached by the combined effect of stellar winds and ionization shocks
originating from the starburst cluster, since models predict a rocket
velocity up to $5~\kms$ \citep{bert89}.

Several studies already suggested the presence of a shell expanding
away from the WR/OB cluster \citep{lisz93, bal01}.  We confirm this
idea but propose that the densest parts of the molecular cloud complex
have a systemic velocity which is given by local Galactic motions and
is not affected by the interaction with the WR/OB cluster.  This
should be particularly true for the vast majority of the compact cloud
fragments identified in Sect.~\ref{s:structure}.  Their detailed
structure and kinematics can, however, be sensitive to the shock wave
propagating ahead of the ionization front.  The kinematics of the W43
molecular cloud complex would merit a careful comparison with models
of ``radiation-driven implosion'' \citep{bert89, lefl94}.  Such a
study will however need further molecular line observations and is
clearly beyond the scope of the present paper.

%% 4.2 The protoclusters
\subsection{Protocluster Candidates in the Mini-starburst W43}\label{s:protoclusters}

We have identified 48 cloud fragments which we think are good sites
for harboring on-going massive star formation.  Their general
characteristics are investigated in Sect.~\ref{s:char} and suggest
they are protoclusters, i.e. molecular clouds that will form clusters
of stars.  We then discuss in Sect.~\ref{s:nature} the evolutionary
state of several of these protoclusters and qualify W43 as a Galactic
mini-starburst in Sect.~\ref{s:starburst}.

%% 4.2.1 General characteristics
\subsubsection{Protocluster candidates}\label{s:char}

In order to determine the nature of the dense, massive, and compact
cloud fragments of W43, we compare their general characteristics with
those of cloud fragments observed in nearby (low- to
intermediate-mass) star-forming regions.  Those cloud fragments are
frequently called ``dense cores'', according to a treshold density
($>10^4~\cmc$) and regardless of their size, mass or substructure
(e.g. \citealt*{jiji99}).  Instead, we prefer to use the terminology
recently established for star-forming molecular clouds like
$\rho$~Ophiuchi or Orion~B when they are surveyed with high-density
and high-resolution tracers such as the submillimeter continuum
(e.g. \citealt{ma01}).  The main advantage is to suggest the level of
subfragmentation expected for any cloud fragment and thus its
abilility to form either one or several stars.  Molecular clouds which
are currently forming clusters of stars are often called
protoclusters, like the $\rho$~Ophiuchi main cloud L1688 or NGC~2068
in Orion~B (e.g. \citealt{wilk83, mott01}).  The best studied case is
the L1688 protocluster which has a mass of $\sim 500~\msun$ in $\sim
1$~pc diameter \citep{wilk83}.  In those protoclusters, starless
condensations are believed to represent the last stage of turbulent
fragmentation ($\sim 0.5~\msun$, $\sim 0.03$~pc), making them the
precursors of individual stars \citep{mott98, mott01}.  Dense cores
are the intermediary cloud structures most probably forming a group of
stars ($\sim 5~\msun$, $\sim 0.1$~pc, e.g. \citealt{jiji99}).

The W43 fragments have diameters ranging from 0.09~pc to 0.56~pc and
masses spanning the range $20~\msun$ to $3\,600~\msun$
(cf. Tables~\ref{t:clumps} and \ref{t:mass}).  On average ($\sim
300~\msun$, $\sim 0.25$~pc), their characteristics are thus closer to
those of protoclusters than those of low-mass dense cores or
condensations.  In that context, most of the cloud fragments observed
in W43 qualify as protocluster candidates.  Higher-resolution studies
are necessary to prove our interpretation but we give below additional
arguments.  Indeed, the \hcops line characteristics of the W43 compact
cloud fragments suggest that they are not homogeneous reservoirs of
gas.  Indeed, optically thin lines are broad ($\Delta v\sim 2-8~\kms$,
cf. Table~\ref{t:lines}) and often non-Gaussian (see
e.g. Figs.~\ref{f:line}a-e).  Moreover, \htcops which is tracing
denser gas than \hcops has narrower lines which sometimes peak at a
slightly different velocity (e.g. W43-MM4, MM7, MM11). \cite{jiji99}
showed that cloud fragments associated with clusters have larger line
widths than those forming single stars.  \cite{myer98} also predicted
the existence of $\sim 0.03$~pc condensations embedded within massive
and turbulent cloud fragments.  Because the compact fragments of W43
are indeed massive ($\sim 300~\msun$) and turbulent ($\Delta v \sim
5~\kms$), they are likely to consist of many smaller-size and denser
structures orbiting within the same cloud.  This interpretation argues
for the protocluster nature of the submillimeter fragments we have
identified.

With present observations only, we cannot draw definite conclusions on
the gravitational status of W43 protoclusters (see
Sect.~\ref{s:virmass}). The small $\msmm/\mvir$ ratios observed for
some of the most massive protocluster candidates suggest the majority
of them could be far from virial equilibrium.  Besides, the mean
ellipticity of the W43 protoclusters (axis ratio 1:2) resembles that
observed for low-mass dense cores and is definitively larger than that
of gravitationally bound starless condensations \citep{mott01}.  It
thus agrees with the idea that W43 protoclusters are generally not in
a gravitation-dominant regime but probably consist of material governed by
turbulence.  However, protocluster candidates are embedded within some
hotter lower-density gas, so that the external pressure could
substantially help confine the protocluster gas and maybe prevent it
from dispersing.  Note that the structures we define here as
protocluster candidates do not need to evolve independently from their
surrounding gas, in which one expects low- to intermediate-mass stars
to form.

Strikingly, these protocluster candidates are very dense (mean density
of $n\htwo\sim 10^6~\cmc$, see Table~\ref{t:mass}) compared to their
low-mass counterparts ( $\sim 10^4~\cmc$, e.g. \citealt{wilk83}).  In
fact they are as dense as the low-mass starless condensations, which
are believed to be the direct progenitors of single (or binary) stars
\citep{mott98}.  Such a dense cloud medium is thus regularly observed
in low-mass star-forming clusters, but not over $\sim 0.25$~pc scale,
which is ten times larger than the typical size of starless
condensations.  Furthermore, we stress that the density of
smaller-scale fragments within those protoclusters will surely be much
higher.  Similar densities, sizes, masses and line widths have already
been recorded for protoclusters harboring UCH\mbox{\sc ~ii}s and HMPOs
\citep{hunt00, plum97, beut02}.  Such cloud characteristics may then
enhance star formation and allow the formation of massive stars.

The unbiased survey we made in the W43 main complex should be
sensitive to protoclusters either being pre-stellar or containing
massive protostars or UC\hii regions.  As all sources are at the same
distance from the Sun and in the same cloud complex, the resulting
sample is far more homogeneous than in any other previous study
(e.g. \citealt{hunt00, srid02}).  No doubt that follow-ups of such a
complete and comprehensive sample will help understand the formation
mechanism of massive stars.  We present below our first attempt to
determine the evolutionary state of W43 protoclusters candidates.

%% 4.2.2 Nature
\subsubsection{Evolutionary state of W43 protoclusters}\label{s:nature}

It is questionable to try to determine the evolutionary state of
protoclusters since, by definition, they contain several YSOs.
However when a high-mass star is forming, it probably shapes the
structure, kinematics, and chemistry of the whole protocluster via
strong heating, ionization and outflow.  It is thus tempting to
constrain the evolution of protoclusters using the evolutionary state
of its main stellar component.  We discuss below the case of 6
protocluster candidates according to their characteristics and the
stellar activity signposts presented in Sect.~\ref{s:mass}.

W43-MM1, MM2 and MM11 are methanol, water (with the exception of MM2)
and hydroxyl maser sources without any infrared or centimeter
detection (cf. Figs.~\ref{f:dust}b, \ref{f:index} and \ref{f:msx}).
W43-MM1 and MM2 are also massive ($\msmm \ga 1\,000~\msun$) luminous
($\lbol \ga 10^4~\lsun$) protoclusters, which are likely to be
gravitationally bound ($\msmm/\mvir >3$).  The gravitational status of
W43-MM11 is less clear, since we measure $\msmm/\mvir \sim 0.35$ (see
Sect.~\ref{s:virmass}).  These protoclusters are globally cold
($\tdust\simeq 20~\K$, e.g. Fig.~\ref{f:sed}a), but recent
observations argue for the presence of embedded ``hot cores'' with
$T\ga 200~\K$ in W43-MM1 and MM2 (Motte et al. in prep.).  The
submillimeter to bolometric luminosity ratio measured in
Sect.~\ref{s:temp} ($\lsmm/\lbol \sim 1.5-3\%$) is similar to that of
low-mass class~0 protostars which are YSOs in their main accretion
phase \citep*{awb00}.  All those rare characteristics make these three
protoclusters excellent candidates for containing massive protostars,
i.e. massive YSOs which have not yet assembled the bulk of their final
stellar mass and neither have developed \hii regions.  As they are not
coincident with any near- to mid-infrared sources, they likely
represent even younger phases than the IRAS selected HMPOs
\citep{bran01, srid02}.  This idea is confirmed by their high mass to
luminosity ratio, sometimes used as an evolutionary indicator.
W43-MM1 and MM2 have higher ratios ($\msmm/\lbol\sim
0.1-0.2~\msun\,\lsun^{-1}$) than those measured\footnote{
   In order to ensure consistency with our derivation, the mass values
   of \cite{beut02} and \cite{hunt00} have been recalculated using
   $\kmm=0.01~\cmg$.}
for HMPOs ($\sim 0.01~\msun\,\lsun^{-1}$, \citealt{srid02}) and
UCH\mbox{\sc ~ii}s envelopes ($\sim 0.006~\msun\,\lsun^{-1}$,
cf. \citealt{hunt00}).

The W43-MM3 protocluster looks very similar to W43-MM1 and MM2 in
terms of mass, luminosity, average temperature and other mass or
luminosity ratios (see Table~\ref{t:mass} and Sect.~\ref{s:temp}).
However, W43-MM3 is not a maser source and coincides with a bright
($S_{\mbox{\tiny 3.5~cm}}\sim 1$~Jy) radio continuum source which is
partially optically thick between 21 and 3.5~cm (see
Fig.~\ref{f:index} and Sect.~\ref{s:crosscorr}).  This compact
free-free emission source is barely resolved, with a $\sim 0.13$~pc
deconvolved diameter smaller than the 0.25~pc size measured at 1.3~mm.
Despite its high luminosity ($\la 10^4~\lsun$), W43-MM3 was not
detected by IRAS, maybe due to confusion from the closeby giant
\hii region.  The inspection of Fig.~\ref{f:msx} suggests that W43-MM3
is loosely associated with MSX G30.7196-0.0854, but this source is
three times more extended than the 1.3~mm protocluster.  Moreover, its
spectral index measured from $8~\micron$ to $21~\micron$ indicates
that the MSX source is not an embedded star/cluster (\citealt{lada87},
see also Fig.~\ref{f:sed}b), but probably lies by chance in front of
the W43-MM3 protocluster.  We thus propose that W43-MM3 harbors an
ultra-compact \hii region, not developed enough to efficiently heat
its $\sim 1\,000~\msun$ of dust and gas.  This source illustrates the
difficulty and inadequacy of the determination of the massive YSOs
evolutionary state from the characteristics of their hosting
protocluster and vice versa.

W43-MM4 is a massive protocluster ($\msmm \sim 500~\msun$), which is
probably gravitationally bound ($\msmm/\mvir \sim 1.4-0.7$,
cf. Sect.~\ref{s:virmass}).  Its study at infrared and centimeter
wavelengths is unfortunately impaired by the strong emission of the
southern ionizing front (see Sect.~\ref{s:giant}).  However in the
21~cm and 3.5~cm surveys of \cite{zoot90} and \cite{beck94}, W43-MM4
coincides with some compact ($\sim 0.09$~pc) free-free emission, which
is strong ($S_{\mbox{\tiny 6~cm}}\sim 300$~mJy) and partially
optically thick, i.e. typical of UC\hii regions.  High-resolution
surveys at centimeter and mid-infrared wavelengths are, however,
necessary to firmly identify W43-MM4 as a protocluster hosting an
UC\hii region.

The W43-MM13 protocluster has a mass of $\sim 200~\msun$ and coincides
with IRAS 18450-0205 (or MSX G30.6882-0.0726) and some free-free
emission peaks detected at 6~cm and 21~cm (see Figs.~\ref{f:msx} and
\ref{f:index}).  The IRAS source is satisfying the criteria
established by \cite{wood89} for selecting UC\hii regions by their
mid- to far-infrared colors (see HIRES flux in
Sect.~\ref{s:crosscorr}).  However, both the MSX and centimeter
sources are extended ({\it FWHM}$\:\sim 1.9$~pc at 21~cm and 0.9~pc at
$21~\micron$), ruling out the interpretation of W43-MM13 containing an
UCH\mbox{\sc ~ii}.  Since the spectral index measured from $8~\micron$
to $60~\micron$ is characteristic of a deeply embedded stellar source
\citep{lada87}, we suggest that a well developed \hii region is
observed behind a foreground cloud.  The visible to mid-infrared
radiation of such an \hii region ($\lbol \sim 10^5~\lsun$) could
easily be extinguished and reddened by the $A_{\rm v} \sim 80$~mag we
measure for the cloud complex containing W43-MM13.  Another
alternative is that the IRAS/MSX and VLA 3~cm/21~cm sources mark a
third ionizing front excited by the WR/OB cluster (see
Sect.~\ref{s:giant}).

We do not have enough information to determine the nature of the
remaining protoclusters of Table~\ref{t:mass}.  Follow-up
observations, searching for ``hot core'' and outflow signatures, as
well as ultracompact free-free emission are in progress.  Mid- to
far-infrared observations with the forthcoming SIRTF and Herschel
satellites will also help in determining their evolutionary state.

%% 4.2.3 Mini-starburst
\subsubsection{The mini-starburst region W43}\label{s:starburst}

As illustrated above, the main molecular complex of W43 looks very
efficient in forming massive stars.  Indeed, at this early stage, we
have already identified five protoclusters containing at least one
massive protostar or UC\hii (Sect.~\ref{s:nature}).  The $\sim 15$
protoclusters in Table~\ref{t:mass}, which have high density
(e.g. $n\htwo \sim 5\times 10^5- 9\times 10^6~\cmc$) and high mass
(e.g. $\msmm \sim 100-3\,600~\msun$) are also good candidates for
being sites of on-going or future massive star formation.
Interestingly, our results preclude that star formation in the W43
main molecular complex is continuous with an efficiency typical of
giant molecular clouds ({\it SFE}$\:\sim 1\%$ over $10^7$~yr,
\citealt{silk97}).  Indeed with a \cite{salp55} initial mass function,
one expects to discover less than one massive YSO in the HMPO or
UC\hii phases (lasting for $10^4-10^5$~yr) when mapping the W43 main
complex.  Therefore, we speculate that star formation is synchronized
within the W43 main molecular complex.  If the $\sim 15$ protoclusters
identified above are indeed currently forming or on the verge of
forming massive stars, the $10^6~\msun$ giant molecular cloud is
likely to disperse within $\sim 10^6$~yr, which is the lifetime of a
massive star.  During this short period, stars should form with an
efficiency that could be as high as {\it SFE}$\:\sim 25\%$, or
equivalently a star formation rate as high as $\sim
0.25~\msun\,$yr$^{-1}$.  Such a star formation efficiency is one order
of magnitude larger than that of normal giant molecular clouds.  It is
similar to that found for the $\rho$~Ophiuchi dense cores, whose size
is one hundred times smaller than the W43 main molecular complex ({\it
SFE}$\:\sim 31\%$ over $10^6$~yr, \citealt{bonte01}).  In $10^6$ years
from now, if W43 becomes a complex of OB clusters encompassed in the
same (14~pc)$^3$ volume, this efficiency would correspond to a stellar
density of $\sim 100$ stars/pc$^3$ (assuming a Salpeter IMF).  A much
larger density has been recorded in the Arches cluster and the core of
the NGC~3603 starburst cluster ($10^5$ stars/pc$^3$,
e.g. \citealt{tapi01, fige99}) but over a $10^4-10^6$ times smaller
volume.  Although this result clearly needs to be confirmed, it
suggests that the W43 molecular complex is experiencing a sudden burst
of star formation.  Therefore, W43, which is known to contain a
starburst cluster of main sequence stars, probably also harbors a
molecular region still actively forming massive stars.  It thus very
likely qualifies as a Galactic mini-starburst region, a term we coin
to define a region whose cloud and stellar content could be used as a
template for those in starburst galaxies.

Since W43 is undergoing a mini-starburst, it offers the unique
opportunity to determine the physical processes responsible for
enhancing star formation in molecular cloud complexes.  Among our
results we show that, despite the extremely high radiation emitted by
the WR/OB cluster, the densest parts of the W43 molecular clouds
remain cool.  Such low temperatures ($\tdust\sim 20~\K$) are unusual
for sites of massive star formation where $\tdust\ga 40~\K$ is
measured \citep{hunt00, srid02}.  Low temperatures coupled with high
densities, could be one key element of the recipe for local
mini-starbursts.  Another obvious element may be the impact of the
revealed cluster of WR/OB stars onto the cloud.  Further studies are
requested before one can make any breakthrough on the origin of such
mini-starburst in our Galaxy.

%%%%%%%%%%%%%%%%%%%%%%%%%%%%%%%%%%%%%%%%%%%%%%%%%%%%%%%%%%%%%%%%%%%%
%% 5 Conclusion
\section{Summary and Conclusion}

W43 is a molecular and \hii complex whose ionized gas has been far
better studied than its molecular component.  We have imaged the W43
main cloud at submillimeter wavelengths in the 1.3~mm and
$350~\micron$ continuum and \hcop(3-2) line emission.  In addition, we
have obtained deep \hcop(3-2) and \htcop(3-2) spectra at selected
locations.  Our main findings can be summarized as follows:

\begin{enumerate}

\item  
  A multiresolution analysis on our submillimeter continuum maps
  identifies $\sim 50$ compact fragments.  Our $350~\micron$ continuum
  observations along with the 3.5~cm image of \cite{bal01} show that
  their 1.3~mm emission is largely thermal emission from cool dust.
\item 
  These bona-fide cloud fragments have diameters varying from 0.09~pc
  to 0.56~pc and masses spanning the range $20~\msun$ to
  $3\,600~\msun$.  Their large size and turbulent line width ($\Delta
  v \sim 5~\kms$) suggest they are protoclusters, i.e. clouds that
  contain many smaller-size and denser structures and will form star
  clusters.  Those protocluster candidates have large mean densities
  ($n\htwo\sim 10^6~\cmc$) reminiscent of the direct progenitors of
  individual low-mass stars.  The W43 protocluster candidates thus
  constitute an excellent sample for studies of the earliest stages of
  massive star formation.
\item
  
  The present, unbiased survey is sensitive to protoclusters either
  being pre-stellar or containing massive protostars or UC\hii
  regions.  Five of the W43 protoclusters are confirmed to contain
  massive YSOs in their HMPO or UC\hii phase.  Follow-up observations
  are needed to determine the evolutionary state of the remaining
  protocluster candidates.
\item
  In W43, the centimeter and infrared emission from the giant \hii
  region dominates, preventing the detection of more compact sources
  that could be associated with UCH\mbox{\sc ~ii}s.  Notably, at least
  two of the three IRAS point sources of the W43 main cloud are not
  stellar in nature; they are instead associated with ionization
  fronts possibly all excited by the closeby WR/OB cluster.
\item
    While the low-density clouds surrounding the starburst cluster may
  have been blown away, the densest parts of the W43 molecular clouds
  seem to remain at the systemic velocity given by Galactic motions.
  A more precise kinematic study is necessary to determine if the
  ionization front associated with the giant \hii region is
  compressing the molecular clouds and triggering star formation.
\item
   W43 is the site of at least two remarkably efficient episodes of
   massive star formation.  Indeed, it is known to harbor a starburst
   cluster containing several WR and OB stars ($\sim 10^6~\lsun$).  We
   show here that the molecular complex is currently undergoing a
   second mini-starburst with a star formation efficiency of $\sim
   25\%/10^6$~yr and possibly a final stellar density of $\sim 100$
   stars/pc$^3$ over (14~pc)$^3$.  Learning about the global
   characteristics of this Galactic mini-starburst region should help
   constraining the physical processes at work in the distant
   starburst galaxies.
\end{enumerate}

\acknowledgments
We are grateful to Dana Balser and Harvey Liszt for the permission to
use their 3.5~cm and 21~cm data and to Carsten Kramer for providing us
with the Gaussclumps program and many helpful comments regarding its
use.  We also thank Philippe Andr\'e, Sylvain Bontemps and the
anonymous referee for useful comments.  The research at the Caltech
Submillimeter Observatory is funded by the NSF through contract
AST-9980846.

%----Tables-----------------------------------------------------
\begin{deluxetable}{lcccccrrc}
\tablecolumns{9} 
\tablewidth{0pc} 
\tabletypesize{\small}
\tablecaption{Compact fragments detected at 1.3~mm and $350~\micron$
\label{t:clumps}}
\tablehead{
\colhead{Fragment} & \multicolumn{2}{c}{Coordinates} & \colhead{\speak} & 
  \colhead{\it FWHM~$^{\rm a}$} & \colhead{\sint} &
  \colhead{\sintsharc~$^{\rm b}$} & \colhead{\spindex} &
  \colhead{\sff/\sint~$^{\rm c}$}\\
\colhead{name} & \colhead{$\alpha_{\rm 2000}$} & 
  \colhead{$\delta_{\rm 2000}$} & \colhead{[\jyb]} &
  \colhead{[pc~$\times$~pc]} & \colhead{[Jy]} & \colhead{[Jy]} &
  \colhead{} & \colhead{[$\%$]}}
\startdata
W43-MM1  & 18:47:47.0 & -1:54:28 & 3.870 & $0.27 \times 0.21$ 
       & 6.490  & 340 & 3.1 & 0\\ 
W43-MM2  & 18:47:36.7 & -2:00:52 & 1.810 & $0.26 \times 0.20$ 
       & 2.920  & 246 & 3.5 & 0$^\star$\\ 
W43-MM3  & 18:47:41.7 & -2:00:26 & 1.070 & $0.27 \times 0.24$ 
       & 1.880  & 140 & 3.4 & 10$^\star$\\ 
W43-MM4  & 18:47:38.3 & -1:57:44 & 0.620 & $0.28 \times 0.22$ 
       & 1.060  & 59 & 3.2 & 10\\ 
W43-MM5  & 18:47:46.3 & -1:54:34 & 0.210 & $0.09 \times 0.09$ 
       & 0.210  & $<18$ & $<3.5$ & 0\\ 
W43-MM6  & 18:47:35.6 & -1:55:16 & 0.290 & $0.51 \times 0.38$ 
       & 0.960  & 65 & 3.3 & 10\\ 
W43-MM7  & 18:47:39.6 & -1:58:34 & 0.330 & $0.76 \times 0.41$ 
       & 1.580  & 170 & 3.7 & 0\\ 
W43-MM8  & 18:47:37.0 & -1:55:28 & 0.290 & $0.79 \times 0.29$ 
       & 1.190  & 33 & 2.6 & 40\\ 
W43-MM9  & 18:47:44.8 & -1:54:43 & 0.310 & $0.32 \times 0.18$ 
       & 0.540  & 56 & 3.6 & 0\\ 
W43-MM10  & 18:47:39.2 & -2:00:30 & 0.280 & $0.29 \times 0.09$ 
       & 0.410  & 23 & 3.2 & 0$^\star$\\ 
W43-MM11  & 18:47:39.7 & -1:57:25 & 0.260 & $0.28 \times 0.09$ 
       & 0.370  & 11 & 2.7 & 10\\ 
W43-MM12  & 18:47:35.9 & -2:01:16 & 0.150 & $0.17 \times 0.09$ 
       & 0.180  & 17 & 3.6 & 10$^\star$\\ 
W43-MM13  & 18:47:36.0 & -2:01:54 & 0.250 & $0.41 \times 0.22$ 
       & 0.530  & 11 & 2.4 & 30$^\star$\\ 
W43-MM14  & 18:47:38.8 & -1:56:49 & 0.240 & $0.42 \times 0.17$ 
       & 0.490  & 16 & 2.7 & 20\\ 
W43-MM15  & 18:47:35.2 & -1:56:36 & 0.240 & $0.46 \times 0.35$ 
       & 0.690  & $<18$ & $<2.6$ & 10\\ 
W43-MM16  & 18:47:40.1 & -1:53:34 & 0.210 & $0.52 \times 0.15$ 
       & 0.470  & $<18$ & $<2.9$ & 0\\ 
W43-MM17  & 18:47:42.5 & -1:59:38 & 0.220 & $0.51 \times 0.24$ 
       & 0.560  & 47 & 3.5 & 0\\ 
W43-MM18  & 18:47:35.6 & -2:02:14 & 0.190 & $0.38 \times 0.11$ 
       & 0.330  & 25 & 3.4 & 0$^\star$\\ 
W43-MM19  & 18:47:33.7 & -1:55:25 & 0.180 & $0.39 \times 0.38$ 
       & 0.490  & $<18$ & $<2.8$ & 40\\ 
W43-MM20  & 18:47:36.4 & -1:56:41 & 0.160 & $0.28 \times 0.16$ 
       & 0.260  & $<18$ & $<3.3$ & 40\\ 
W43-MM21  & 18:47:39.3 & -1:53:53 & 0.160 & $0.50 \times 0.22$ 
       & 0.390  & 65 & 4.0 & 0\\ 
W43-MM22  & 18:47:27.0 & -2:00:36 & 0.170 & $0.71 \times 0.26$ 
       & 0.610  & 31 & 3.1 & 10$^\star$\\ 
W43-MM23  & 18:47:41.7 & -1:53:23 & 0.150 & $0.35 \times 0.13$ 
       & 0.250  & $<18$ & $<3.3$ & 0\\ 
W43-MM24  & 18:47:55.9 & -1:53:30 & 0.160 & $0.56 \times 0.47$ 
       & 0.670  & $<18$ & $<2.6$ & 0$^\star$\\ 
W43-MM25  & 18:47:27.6 & -1:56:55 & 0.110 & $0.31 \times 0.09$ 
       & 0.170  & $<18$ & $<3.7$ & 10\\ 
W43-MM26  & 18:47:40.8 & -1:54:21 & 0.150 & $0.46 \times 0.31$ 
       & 0.410  & 37 & 3.5 & 0\\ 
W43-MM27  & 18:47:42.6 & -1:56:17 & 0.130 & $0.28 \times 0.14$ 
       & 0.210  & 15 & 3.4 & 0\\ 
W43-MM28  & 18:47:33.7 & -2:01:03 & 0.130 & $0.33 \times 0.09$ 
       & 0.190  & $<18$ & $<3.6$ & 0$^\star$\\ 
W43-MM29  & 18:47:52.4 & -1:54:57 & 0.140 & $0.25 \times 0.09$ 
       & 0.180  & $<18$ & $<3.6$ & 10$^\star$\\ 
W43-MM30  & 18:47:30.2 & -2:00:46 & 0.140 & $0.57 \times 0.43$ 
       & 0.560  & 22 & 2.9 & 0$^\star$\\ 
W43-MM31  & 18:47:40.5 & -1:58:52 & 0.120 & $0.42 \times 0.16$ 
       & 0.240  & 14 & 3.2 & 0\\ 
W43-MM32  & 18:47:38.5 & -1:55:28 & 0.120 & $0.35 \times 0.18$ 
       & 0.210  & 12 & 3.2 & 70\\ 
W43-MM33  & 18:47:50.2 & -1:54:04 & 0.110 & $0.41 \times 0.09$ 
       & 0.200  & $<18$ & $<3.5$ & 0$^\star$\\ 
W43-MM34  & 18:47:27.9 & -2:01:19 & 0.090 & $0.31 \times 0.11$ 
       & 0.150  & $<18$ & $<3.8$ & 10$^\star$\\ 
W43-MM35  & 18:47:32.9 & -1:57:14 & 0.100 & $0.76 \times 0.23$ 
       & 0.340  & $<18$ & $<3.1$ & 0\\ 
W43-MM36  & 18:47:27.1 & -1:57:43 & 0.090 & $0.24 \times 0.09$ 
       & 0.120  & $<18$ & $<3.9$ & 0\\ 
W43-MM37  & 18:47:27.6 & -1:56:36 & 0.110 & $0.23 \times 0.09$ 
       & 0.140  & $<18$ & $<3.8$ & 10\\ 
W43-MM38  & 18:47:24.3 & -1:55:59 & 0.100 & $0.74 \times 0.40$ 
       & 0.450  & $<18$ & $<2.9$ & 10\\ 
W43-MM39  & 18:47:42.3 & -1:52:11 & 0.090 & $0.63 \times 0.13$ 
       & 0.230  & $<18$ & $<3.4$ & 0$^\star$\\ 
W43-MM40  & 18:47:26.6 & -1:57:22 & 0.090 & $0.40 \times 0.21$ 
       & 0.190  & $<18$ & $<3.6$ & 0\\ 
W43-MM41  & 18:47:25.3 & -1:56:55 & 0.090 & $0.50 \times 0.37$ 
       & 0.270  & $<18$ & $<3.3$ & 10\\ 
W43-MM42  & 18:47:41.3 & -1:59:23 & 0.070 & $0.35 \times 0.09$ 
       & 0.110  & $<18$ & $<4.0$ & 0\\ 
W43-MM43  & 18:47:27.4 & -1:56:08 & 0.080 & $0.51 \times 0.09$ 
       & 0.170  & $<18$ & $<3.7$ & 10\\ 
W43-MM44  & 18:47:36.4 & -1:58:17 & 0.080 & $0.39 \times 0.26$ 
       & 0.170  & 22 & 3.8 & 0\\ 
W43-MM45  & 18:47:34.4 & -1:57:02 & 0.070 & $0.09 \times 0.09$ 
       & 0.070  & $<18$ & $<4.4$ & 0\\ 
W43-MM46  & 18:47:56.8 & -1:55:02 & 0.070 & $0.09 \times 0.09$ 
       & 0.070  & $<18$ & $<4.4$ & 10$^\star$\\ 
W43-MM47  & 18:47:30.9 & -1:57:41 & 0.070 & $0.42 \times 0.16$ 
       & 0.140  & $<18$ & $<3.8$ & 0\\ 
W43-MM48  & 18:47:38.1 & -1:59:10 & 0.070 & $0.42 \times 0.09$ 
       & 0.120  & $<18$ & $<3.9$ & 0\\ 
W43-MM49  & 18:47:47.9 & -1:54:01 & 0.090 & $0.09 \times 0.09$ 
       & 0.090  & $<18$ & $<4.2$ & 0\\ 
W43-MM50  & 18:47:25.1 & -1:55:24 & 0.060 & $0.65 \times 0.18$ 
       & 0.170  & $<18$ & $<3.6$ & 20\\ 
W43-MM51  & 18:47:34.7 & -2:01:02 & 0.070 & $0.09 \times 0.09$ 
       & 0.070  & $<18$ & $<4.4$ & $<10$$^\star$\\ 
\enddata
\tablenotetext{a}{Deconvolved {\it FWHM} size derived from a
2D-Gaussian fit to the 1.3~mm map (after background subtraction,
cf. Sect.~\ref{s:structure}).  An upper limit {\it FWHM} of 0.09~pc
was assumed for unresolved fragments.}
\tablenotetext{b}{Integrated flux measured at $350~\micron$ (after
background subtraction).  Upper limits of $2\times 3\sigma$ are given
when the 1.3~mm fragment is undetected at $350~\micron$.  The factor 2
accounts for the average $\sint/\speak$ ratio of these compact but
resolved sources.}
\tablenotetext{c}{Free-free contamination of the 1.3~mm fluxes
estimated from the 3.5~cm map of \citet{bal01} or the 21~cm map of
\citet{lisz93} when indicated by a star maker.}
\end{deluxetable}

\begin{deluxetable}{llcrcrc}
\tablecolumns{7} 
\tablewidth{0pc}
\tablecaption{\hcops line survey at selected locations\label{t:lines}}
\tablehead{
\colhead{Fragment} & \colhead{Molecular} & \colhead{Component} & 
  \colhead{\vlsr} & \colhead{\tmb} & \colhead{$\int \tmb~dv$} &
  \colhead{$\Delta v$}\\
\colhead{name} & \colhead{transition} & \colhead{number} &
  \colhead{[\kms]} & \colhead{[K]} & \colhead{[K \kms]} &
  \colhead{[\kms]}}
\startdata
W43-MM1 & \htcop(3-2) & 1 & 98.8 & $1.30 \pm 0.10$ & 8.2 & 5.9\\
W43-MM2 & \htcop(3-2) & 1 & 90.8 & $1.04 \pm 0.09$ & 4.6 & 4.2\\
W43-MM3 & \htcop(3-2) & 1 & 93.5 & $0.82 \pm 0.06$ & 3.8 & 4.4\\
W43-MM4 & \hcop(3-2) & 1 & 91.8 & $3.24 \pm 0.14$ & 17.5 & 5.1\\
        &            & 2 & 97.6 & $1.83 \pm 0.14$ & 6.0 & 3.1\\
        & \htcop(3-2) & 1 & 92.0 & $1.55 \pm 0.13$ & 4.4 & 2.7\\
        &            & 2 & 97.3 & $0.20 \pm 0.13$ & 0.7 & 3.2\\
W43-MM6\&8 & \hcop(3-2) & 1 & 94.4 & $2.37 \pm 0.14$ & 20.9 & 8.3\\
        & \htcop(3-2) & 1 & 94.3 & $0.35 \pm 0.06$ & 2.4 & 6.3\\
W43-MM7 & \hcop(3-2) & 1 & 90.6 & $0.83 \pm 0.14$ & 3.9 & 4.4\\
        &            & 2 & 97.4 & $1.72 \pm 0.14$ & 6.7 & 3.7\\
        & \htcop(3-2) & 1 & 91.9 & $0.19 \pm 0.06$ & 0.7 & 3.7\\
        &            & 2 & 96.0 & $0.31 \pm 0.06$ & 0.7 & 2.2\\
W43-MM9 & \htcop(3-2) & 1 & 96.0 & $0.61 \pm 0.11$ & 3.6 & 5.5\\
W43-MM10 & \htcop(3-2) & 1 & 92.9 & $0.59 \pm 0.11$ & 2.3 & 3.7\\
W43-MM11 & \hcop(3-2) & 1 & 91.5 & $2.02 \pm 0.10$ & 14.4 & 6.7\\
        & \htcop(3-2) & 1 & 91.8 & $0.80 \pm 0.13$ & 3.2 & 3.8\\
W43-MM14 & \hcop(3-2) & 1 & 89.9 & $1.01 \pm 0.08$ & 6.6 & 6.1\\
W43-MM15\&20 & \hcop(3-2) & 1 & 81.0 & $0.48 \pm 0.05$ & 2.4 & 4.8\\
        &            & 2 & 87.0 & $0.23 \pm 0.05$ & 0.9 & 3.5\\
        &            & 3 & 94.4 & $0.58 \pm 0.05$ & 4.2 & 6.8\\
        &            & 4 & 107.7 & $0.10 \pm 0.05$ & 0.7 & 6.0\\
W43-MM25 & \hcop(3-2) & 1 & 92.3 & $0.67 \pm 0.09$ & 3.5 & 4.9\\
\enddata
\end{deluxetable}

\begin{deluxetable}{lcrcrl}
\tablecolumns{5} 
\tablewidth{0pc} 
\tablecaption{Properties of submillimeter compact fragments\label{t:mass}}
\tablehead{
\colhead{Fragment} & \colhead{\tdust} & \colhead{\msmm} &
  \colhead{$<n\htwo>$$^a$} & \colhead{nb$\:\times\mvir$$~^b$}\\
\colhead{name} & \colhead{[K]} & \colhead{[\msun]} & 
  \colhead{[\cmc]} & \colhead{[\msun]} }
\startdata
\noalign{\smallskip}
W43-MM1 & 20 & 3590 & $8.8  \times 10^6$ & 1040 \\ 
W43-MM2 & 20 & 1620 & $4.6  \times 10^6$ & 490 \\ 
W43-MM3 & 20 & 960 & $2.0  \times 10^6$ & 610 \\ 
W43-MM4 & 20 & 520 & $1.1  \times 10^6$ & $2\times 370$ \\ 
W43-MM5 & 20 & 110 & $5.3  \times 10^6$ & \nodata \\ 
W43-MM6 & 20 & 500 & $2.0  \times 10^5$ & 2930$^\star$ \\ 
W43-MM7 & 20 & 870 & $1.7  \times 10^5$ & $2\times 840$ \\ 
W43-MM8 & 20 & 390 & $1.2  \times 10^5$ & 2930$^\star$ \\ 
W43-MM9 & 20 & 290 & $6.6  \times 10^5$ & 920 \\
W43-MM10 & 20 & 230 & $1.8  \times 10^6$ & 270 \\
W43-MM11 & 20 & 190 & $1.6  \times 10^6$ & 530 \\ 
W43-MM12 & 20 & 90 & $1.7  \times 10^6$ & \nodata \\ 
W43-MM13 & 20 & 200 & $2.5  \times 10^5$ & \nodata \\ 
W43-MM14 & 20 & 230 & $3.8  \times 10^5$ & 1260 \\
W43-MM15 & 30 & 200 & $1.1  \times 10^5$ & $4\times 1380^\star$ \\
W43-MM16 & 20 & 260 & $4.2  \times 10^5$ & \nodata \\ 
W43-MM17 & 20 & 310 & $2.5  \times 10^5$ & \nodata \\ 
W43-MM18 & 20 & 180 & $7.5  \times 10^5$ & \nodata \\ 
W43-MM19 & 20 & 180 & $1.0  \times 10^5$ & \nodata \\ 
W43-MM20 & 30 & 50 & $1.9  \times 10^5$ & $4\times 1380^\star$ \\ 
W43-MM21 & 20 & 210 & $2.0  \times 10^5$ & \nodata \\ 
W43-MM22 & 20 & 320 & $1.3  \times 10^5$ & \nodata \\ 
W43-MM23 & 20 & 140 & $4.9  \times 10^5$ & \nodata \\ 
W43-MM24 & 20 & 370 & $9.1  \times 10^4$ & \nodata \\ 
W43-MM25 & 30 & 50 & $3.5  \times 10^5$ & 490 \\ 
W43-MM26 & 20 & 230 & $1.4  \times 10^5$ & \nodata \\ 
W43-MM27 & 20 & 110 & $4.8  \times 10^5$ & \nodata \\ 
W43-MM28 & 20 & 110 & $6.9  \times 10^5$ & \nodata \\ 
W43-MM29 & 20 & 90 & $9.3  \times 10^5$ & \nodata \\ 
W43-MM30 & 20 & 300 & $8.3  \times 10^4$ & \nodata \\ 
W43-MM31 & 20 & 130 & $2.6  \times 10^5$ & \nodata \\ 
W43-MM33 & 20 & 110 & $5.1  \times 10^5$ & \nodata \\ 
W43-MM34 & 20 & 70 & $4.0  \times 10^5$ & \nodata \\ 
W43-MM35 & 30 & 110 & $5.2  \times 10^4$ & \nodata \\ 
W43-MM36 & 30 & 40 & $3.9  \times 10^5$ & \nodata \\ 
W43-MM37 & 30 & 40 & $4.5  \times 10^5$ & \nodata \\ 
W43-MM38 & 30 & 140 & $2.8  \times 10^4$ & \nodata \\ 
W43-MM39 & 20 & 120 & $1.8  \times 10^5$ & \nodata \\ 
W43-MM40 & 30 & 60 & $8.5  \times 10^4$ & \nodata \\ 
W43-MM41 & 30 & 80 & $3.4  \times 10^4$ & \nodata \\ 
W43-MM42 & 20 & 60 & $3.9  \times 10^5$ & \nodata \\ 
W43-MM43 & 30 & 50 & $1.7  \times 10^5$ & \nodata \\ 
W43-MM44 & 20 & 100 & $1.0  \times 10^5$ & \nodata \\ 
W43-MM45 & 30 & 20 & $1.0  \times 10^6$ & \nodata \\ 
W43-MM47 & 30 & 50 & $8.9  \times 10^4$ & \nodata \\ 
W43-MM48 & 20 & 60 & $3.0  \times 10^5$ & \nodata \\ 
W43-MM49 & 20 & 50 & $2.2  \times 10^6$ & \nodata \\ 
W43-MM51 & 20 & 40 & $1.8  \times 10^6$ & \nodata \\ 
\enddata
\tablenotetext{a}{Mean density derived from Col.~3, and Col.~5 of
Table~\ref{t:clumps}: $<n\htwo> = \msmm / \left[
\frac{4}{3}\pi\times(\mbox{\rm \it FWHM}/2)^3 \right]$.}
\tablenotetext{b}{Virial mass of an average velocity component
multiplied by the number of components.  Virial masses that correspond
to two independent submillimeter fragments are indicated by a star
marker.}
\end{deluxetable}

%---------Figures----------
\begin{figure*}
\includegraphics[width=12.2cm,angle=270]{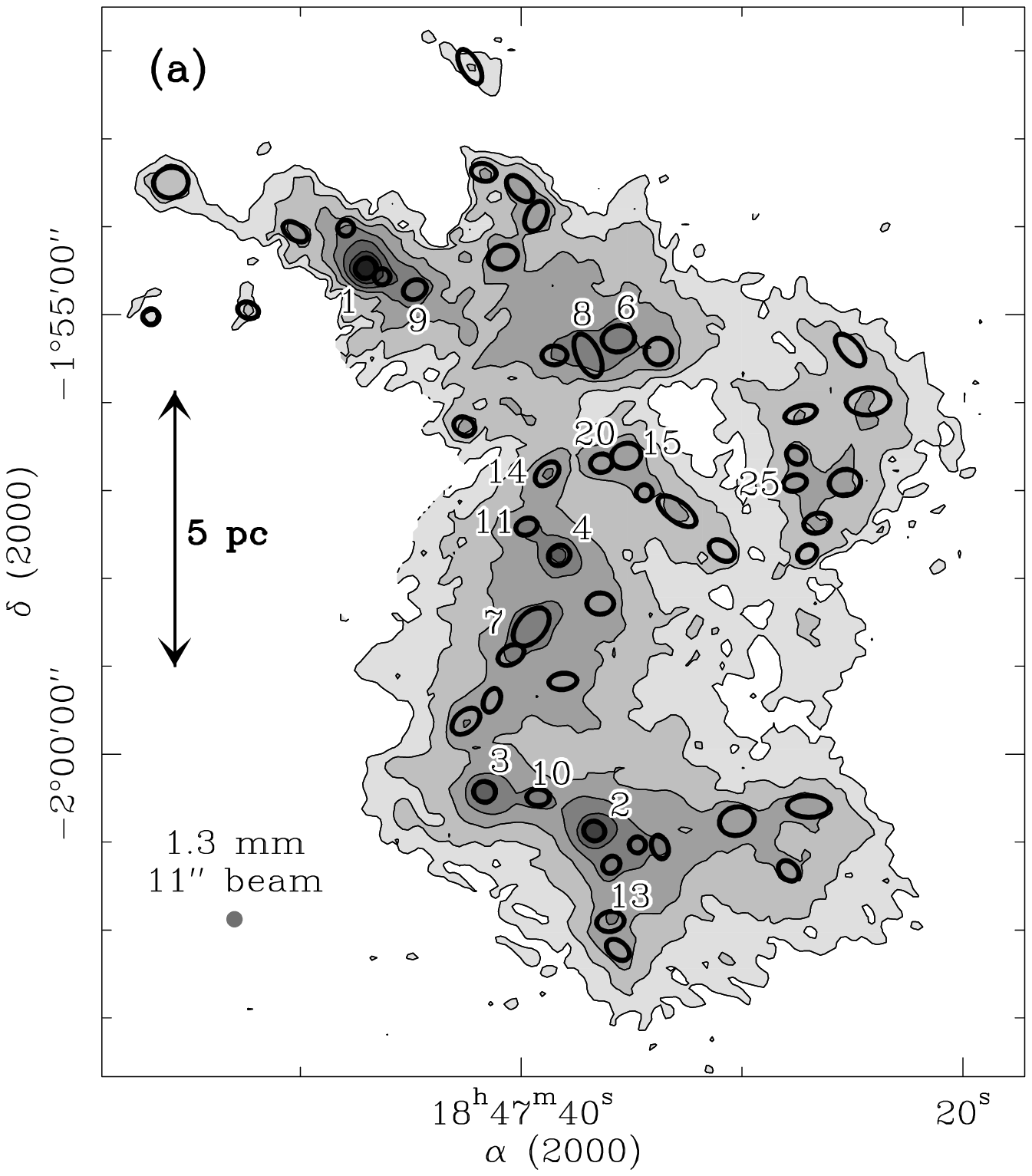}
\hskip -16.8cm
\includegraphics[width=12.2cm,angle=270]{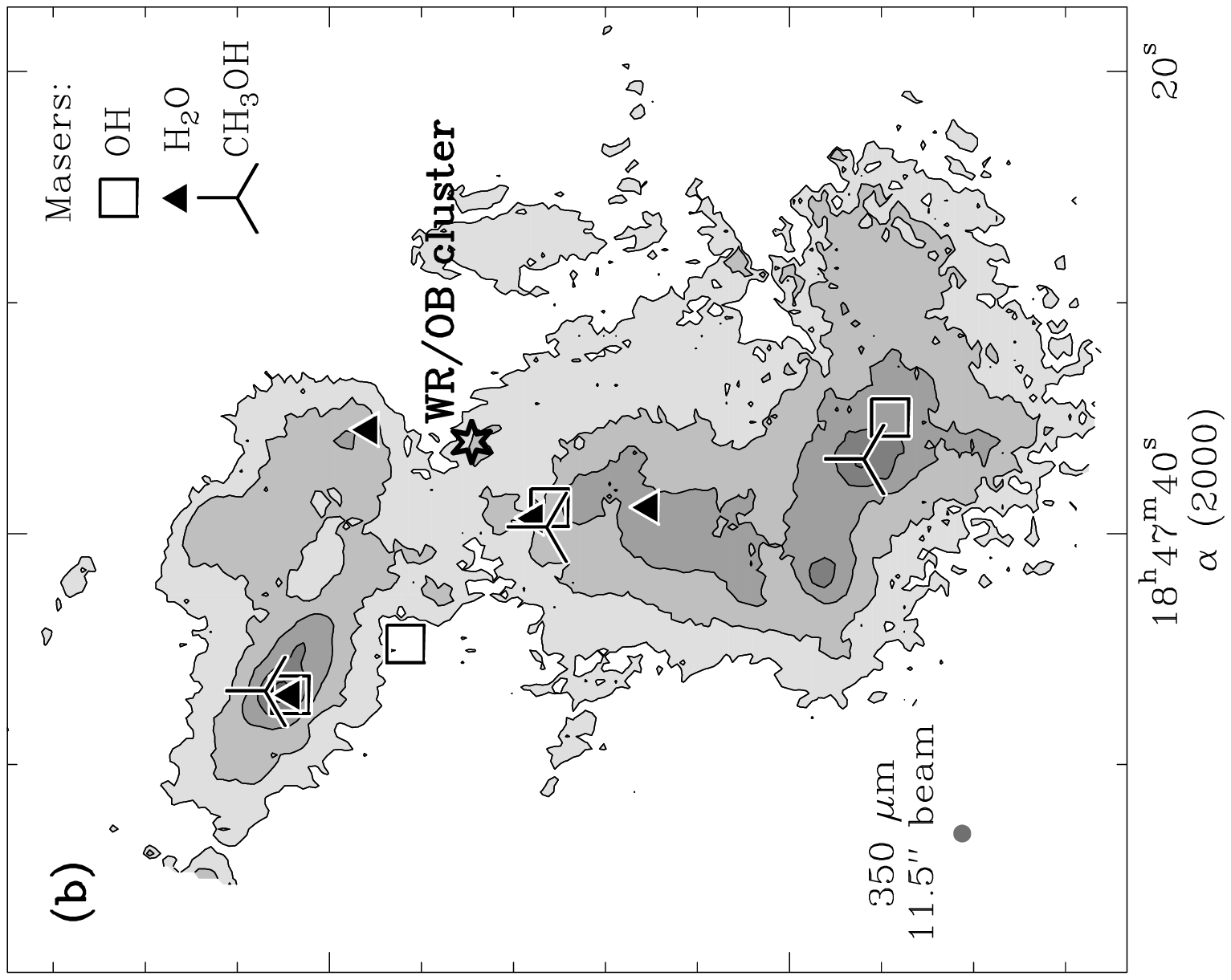}
\vskip -0.5cm
\caption[]{The W43 main star-forming complex mapped at 1.3~mm with
MAMBO ({\bf a}) and $350~\micron$ with SHARC ({\bf b}).  Contour
levels are logarithmic, spaced by factors 2.  Levels and rms noise
are: {\bf (a)}~60, 120, 240, 480, 960~m\jyb, and 2, 4~\jyb,
$1\sigma\simeq 12$~mJy/11\arcsec-beam; {\bf (b)}~12, 24, 48, 96 and
200~\jyb, $1\sigma\simeq 3$~Jy/11.5\arcsec-beam.  In {\bf (a)}, the
Gaussian dust fragments extracted from the 1.3~mm map are indicated by
ellipses, fragments which are relevant for the discussion are also
labelled.  In {\bf (b)}, the WR/OB association is indicated by a star
symbol.  OH, H$_2$O and CH$_3$OH maser sources are also shown as open
squares, filled triangles, and three branches crosses, respectively.}
\label{f:dust}
\end{figure*}

\begin{figure*}
\includegraphics[width=9cm,angle=270]{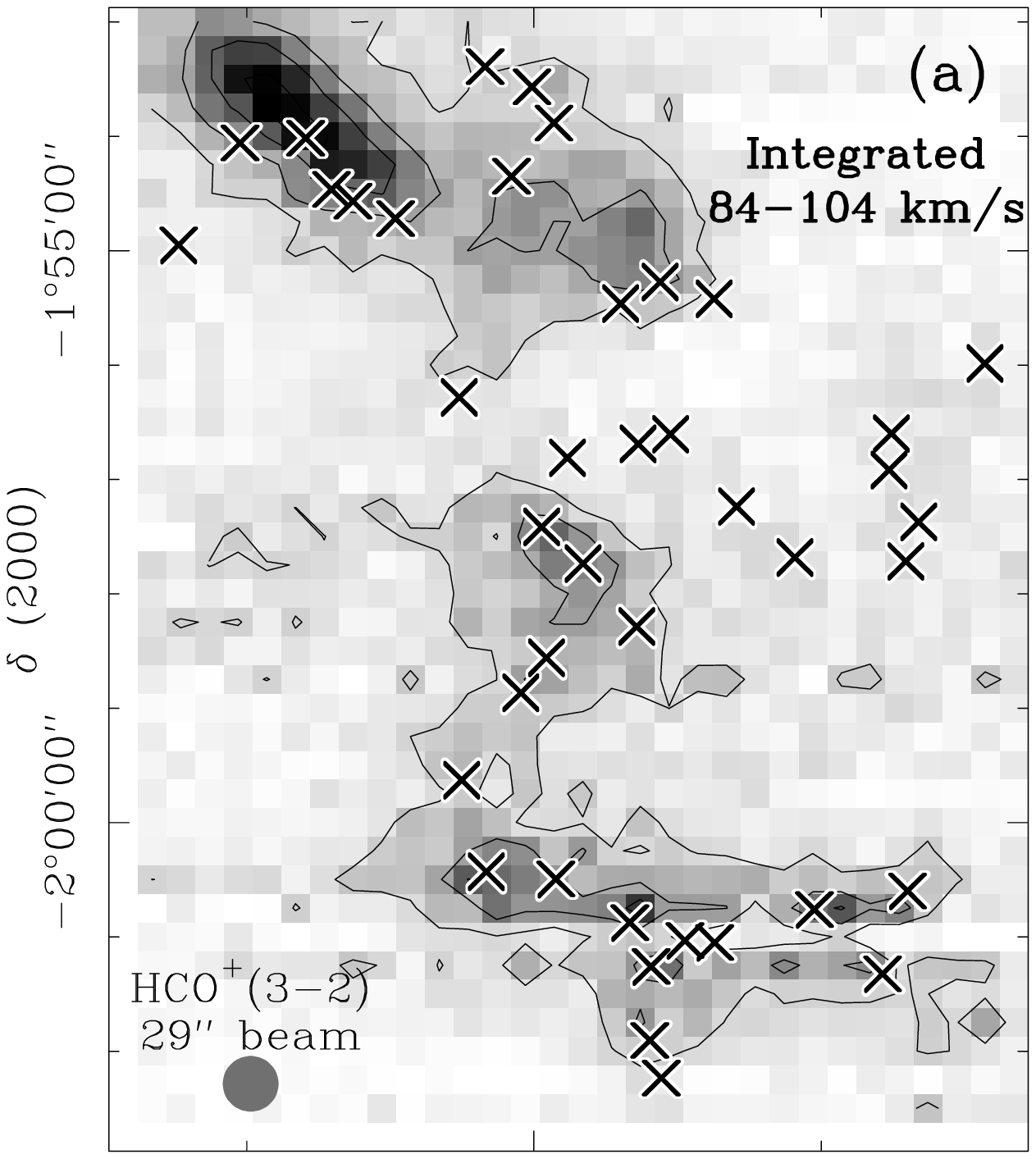}
\hskip -13.3cm
\includegraphics[width=9cm,angle=270]{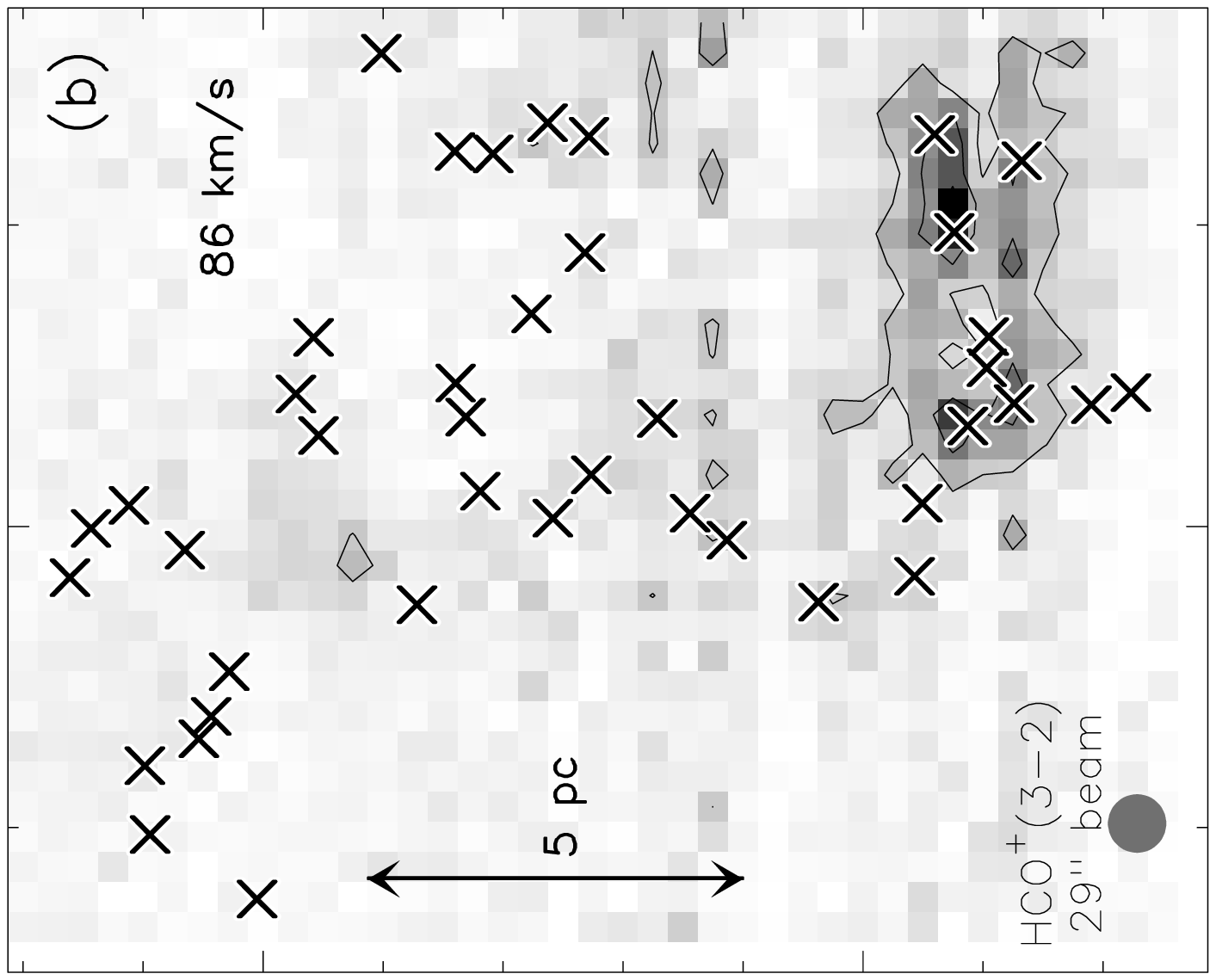}
\hskip -7.05cm
\includegraphics[width=9cm,angle=270]{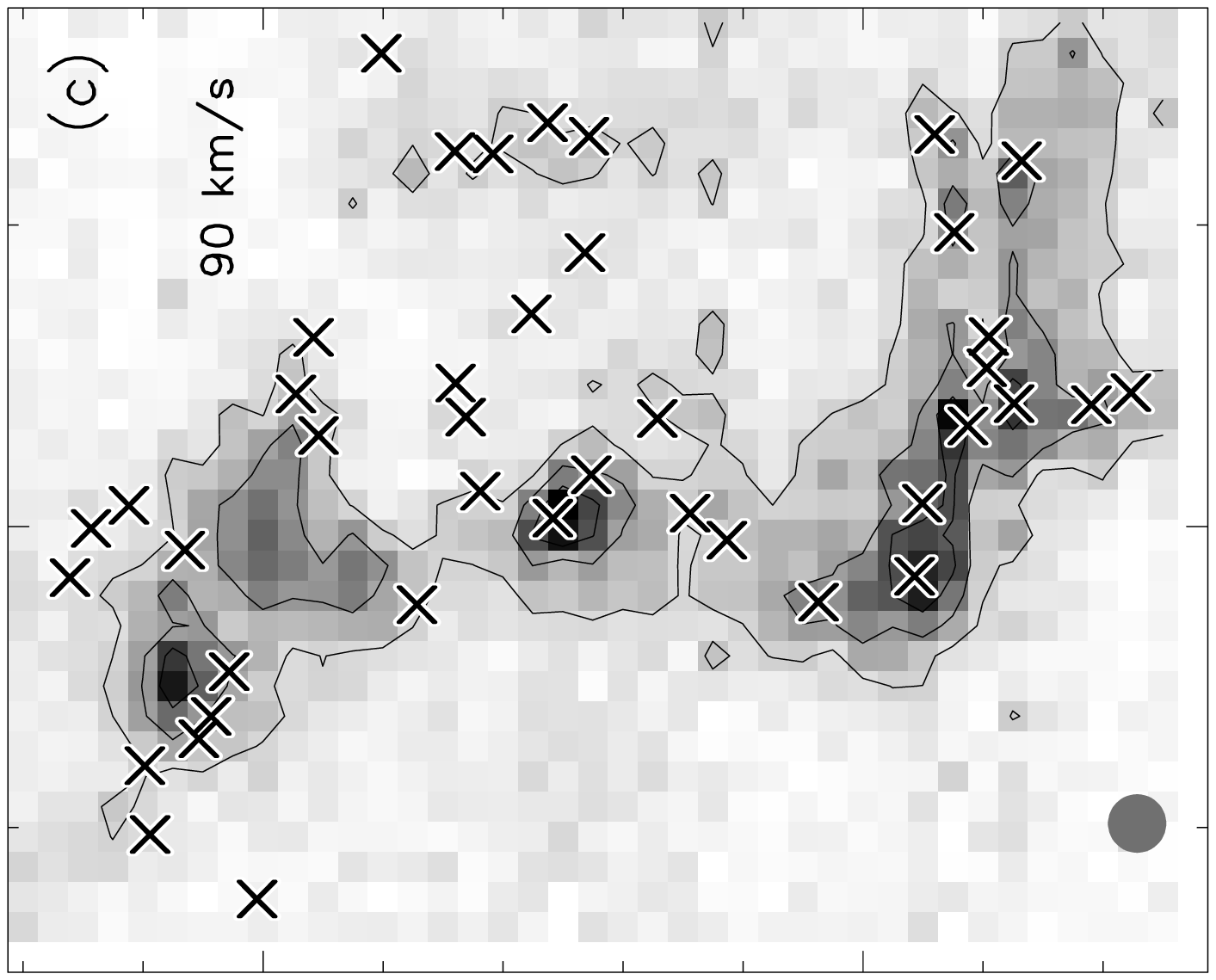}
\vskip -1.7cm
\includegraphics[width=9cm,angle=270]{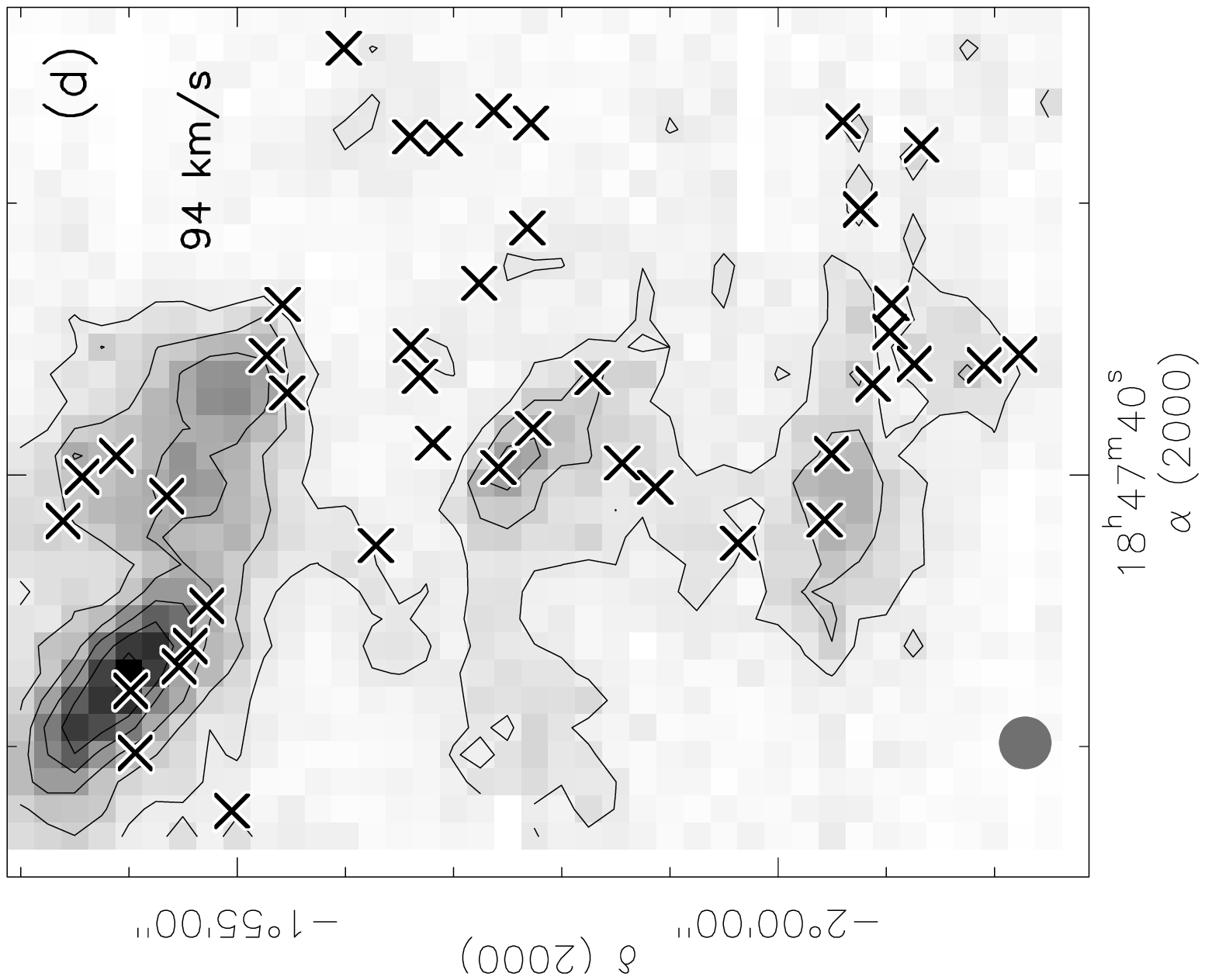}
\hskip -13.3cm
\includegraphics[width=9cm,angle=270]{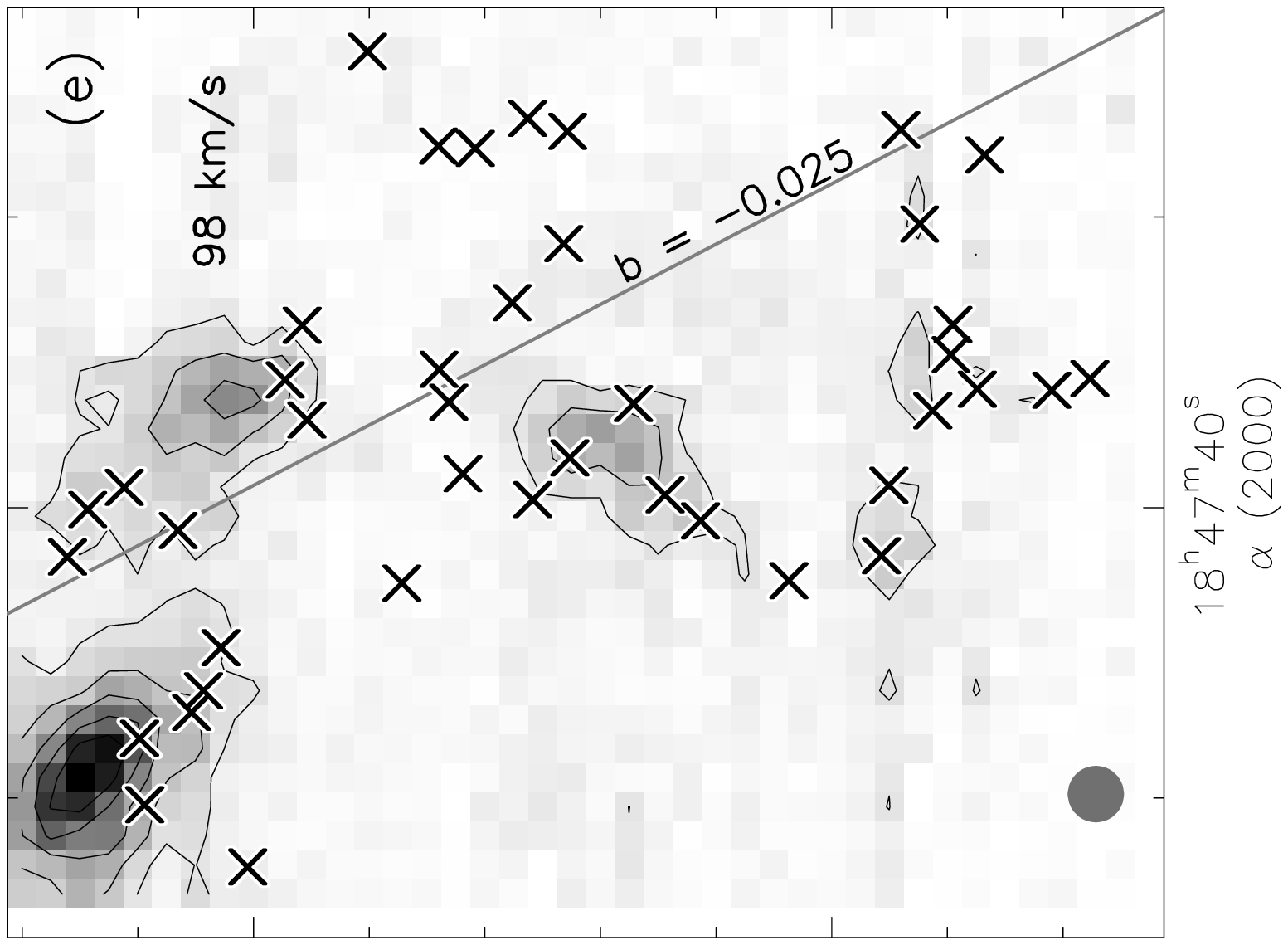}
\hskip -7.05cm
\includegraphics[width=9cm,angle=270]{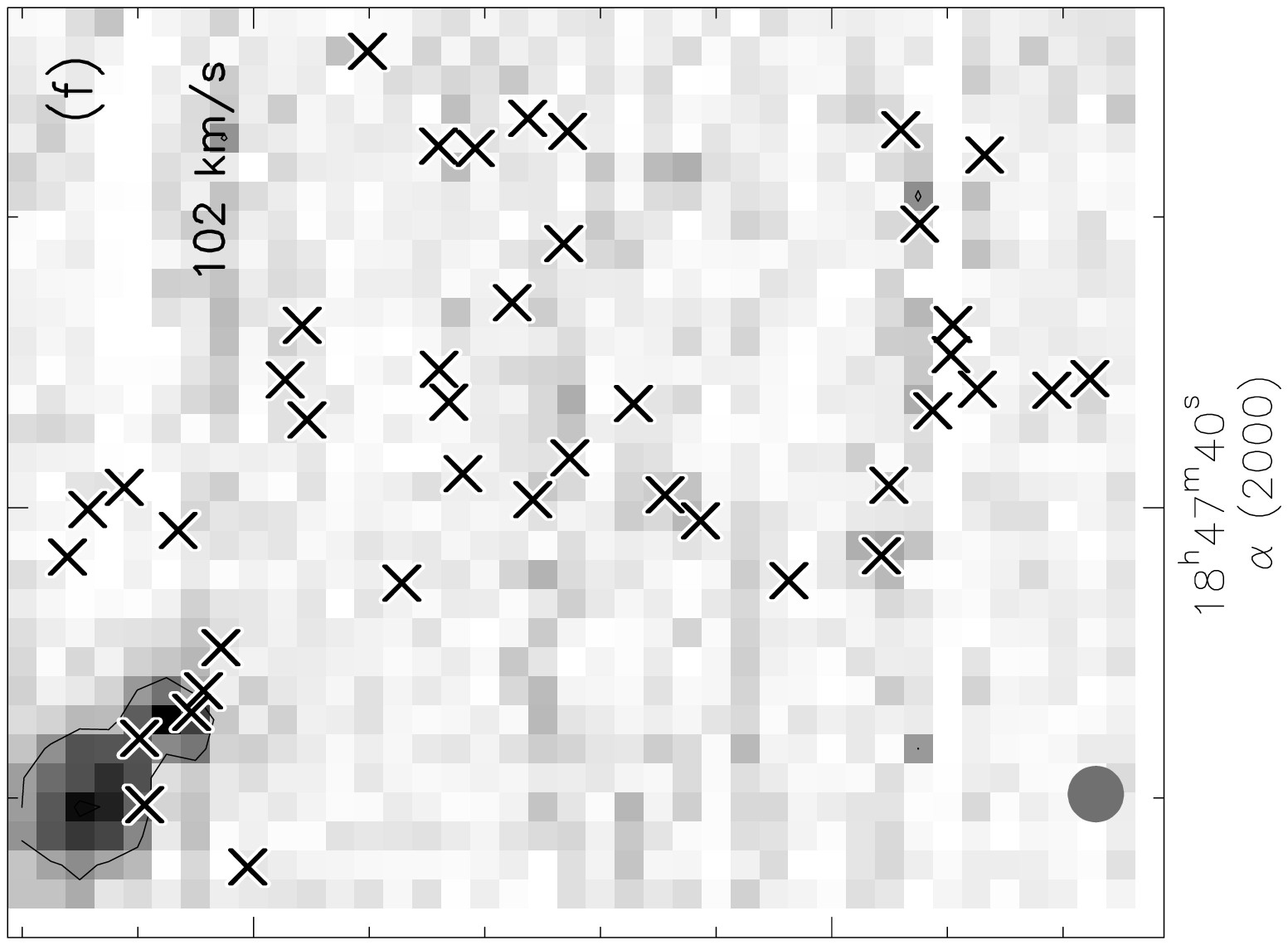}
\vskip -0.3cm
\caption[]{The W43 main star-forming complex mapped in \hcop(3-2).
In {\bf (a)}, the intensity map is integrated over the $84-104~\kms$
velocity range.  Contour levels are 9 to 36 by $9~\K\,\kms$ and rms
noise is $1\sigma\sim 3~\K\,\kms$.  In {\bf (b)}-{\bf (f)}, the
channel maps are integrated over 4~\kms channels.  Contour levels are
0.7 to 4.2 by $0.7~\K\,\kms$ and rms noise is $1\sigma\sim
0.23~\K\,\kms$.  The positions of dust fragments extracted from
Fig.~\ref{f:dust}a are marked by crosses.  The Galactic plane at the
WR/OB cluster latitude is displayed in {\bf (e)}.}
\label{f:hcop}
\end{figure*}

\begin{figure}
\includegraphics[width=11.5cm,angle=270]{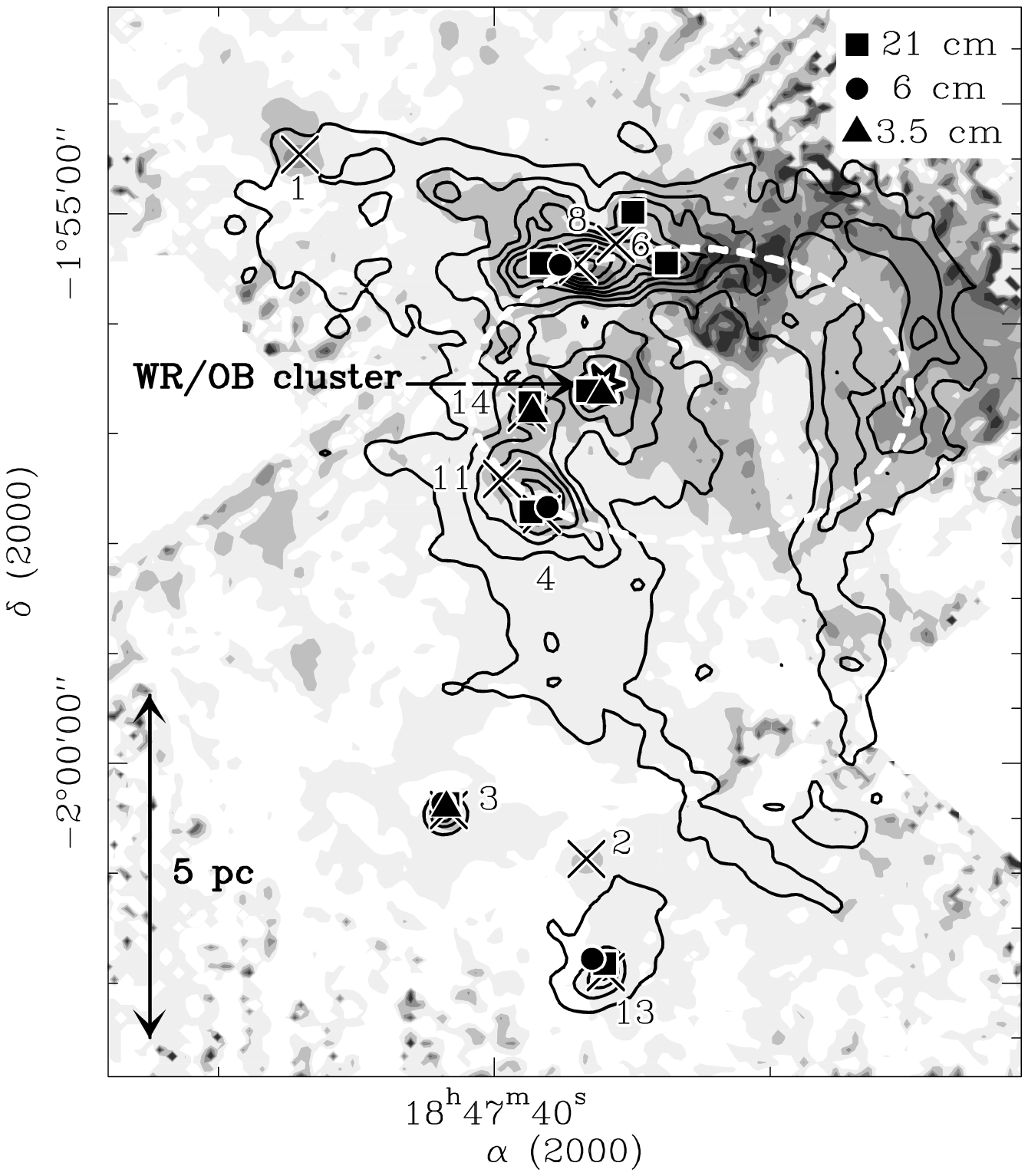}
\vskip -0.5cm
\caption[]{The spectral index map of W43 measured between
$350~\micron$ and 1.3~mm (gray scale) compared with the free-free
emission mapped at 21~cm with the VLA (contours,
cf. \citealt{lisz93}).  The submillimeter spectral index $\spindex$
varies from 2 (black level) to 4 (white level) with a step of 0.5.
Contour levels are $10\%$ to $100\%$ by $10\%$ with a maximum
centimeter flux of $\sim 0.8$~Jy/12.5\arcsec-beam.  The WR/OB
association is indicated by the star symbol, the ionization front
caused by its UV radiation is outlined by a thick dashed white
ellipse.  Small-diameter sources identified at 21~cm, 6~cm and 3.5~cm
wavelengths are marked by filled squares, circles, and triangles,
respectively.  Selected submillimeter fragments are indicated with
crosses.}
\label{f:index}
\end{figure}

\begin{figure}
\includegraphics[width=11cm,angle=0]{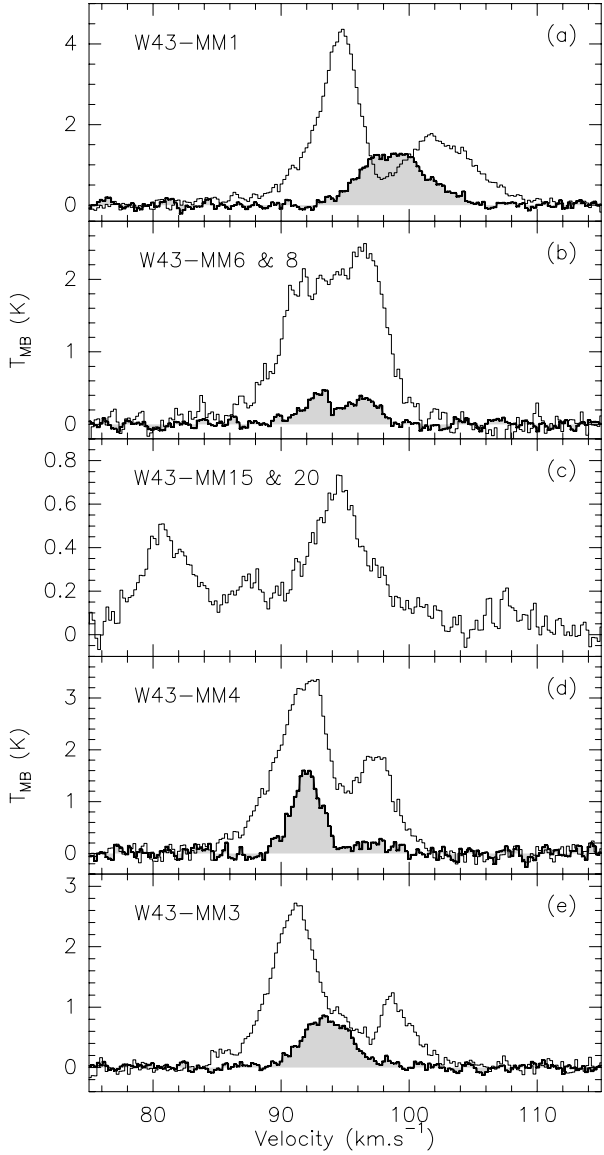}
\vskip -0.2cm
\caption[]{\hcop(3-2) (thin line histogram) and \htcop(3-2) (thick
line histogram) spectra smoothed to a resolution of $0.6~\kms$.  In
{\bf (a)}-{\bf (b)} and {\bf (d)}-{\bf (e)} the sources are located
north (resp. south) of the WR/OB cluster which lies at location
W43-MM15\&20.}
\label{f:line}
\end{figure}

\begin{figure}
\includegraphics[width=9cm,angle=0]{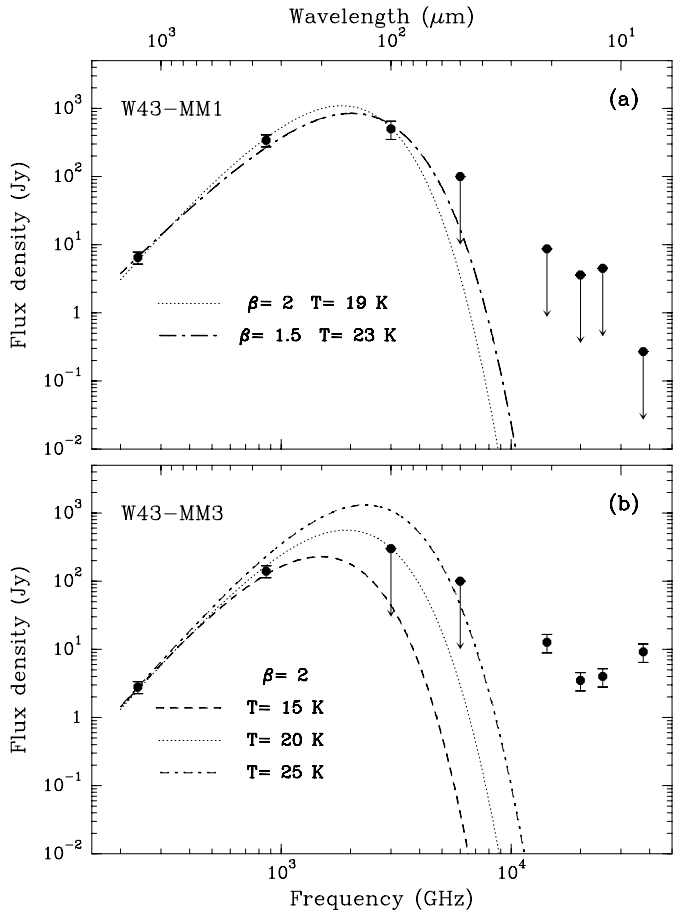}
\vskip -0.2cm
\caption[]{Spectral energy distributions of the cloud fragments W43-MM1
({\bf a}) and W43-MM3 ({\bf b}) compared with gaybody models with
varying dust opacity indices and temperatures.  The absolute
uncertainty of MSX and KAO fluxes is set to $30\%$, that of MAMBO and
SHARC fluxes is $20\%$.}
\label{f:sed}
\end{figure}

\begin{figure}
\includegraphics[width=11.5cm,angle=270]{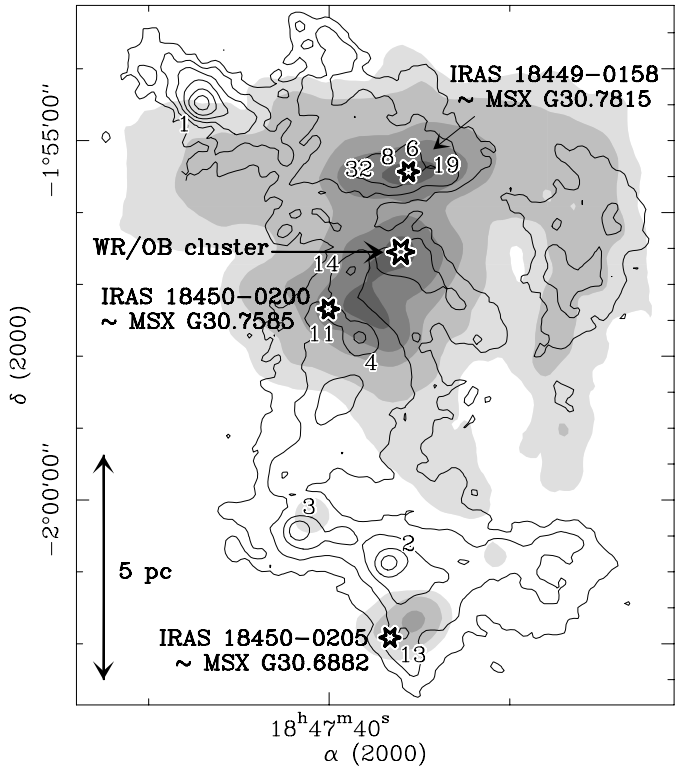}
\vskip -0.5cm
\caption[]{The continuum emission of W43 at $21~\micron$ (gray
scale) compared with that at 1.3~mm (contours).  The $21~\micron$ map
was obtained by the MSX satellite with a $20\arcsec$ aperture.  Levels
are 0.5, 1, 3, 6, and 12 $\times 10^9$~Jy$\,$sr$^{-1}$. The 1.3~mm map
is shown in Fig.~\ref{f:dust}a with the same contours, except the
first one.  IRAS/MSX sources and the WR/OB association are indicated
by star symbols; selected submillimeter fragments are labelled.}
\label{f:msx}
\end{figure}

\begin{figure}
\includegraphics[width=9cm,angle=0]{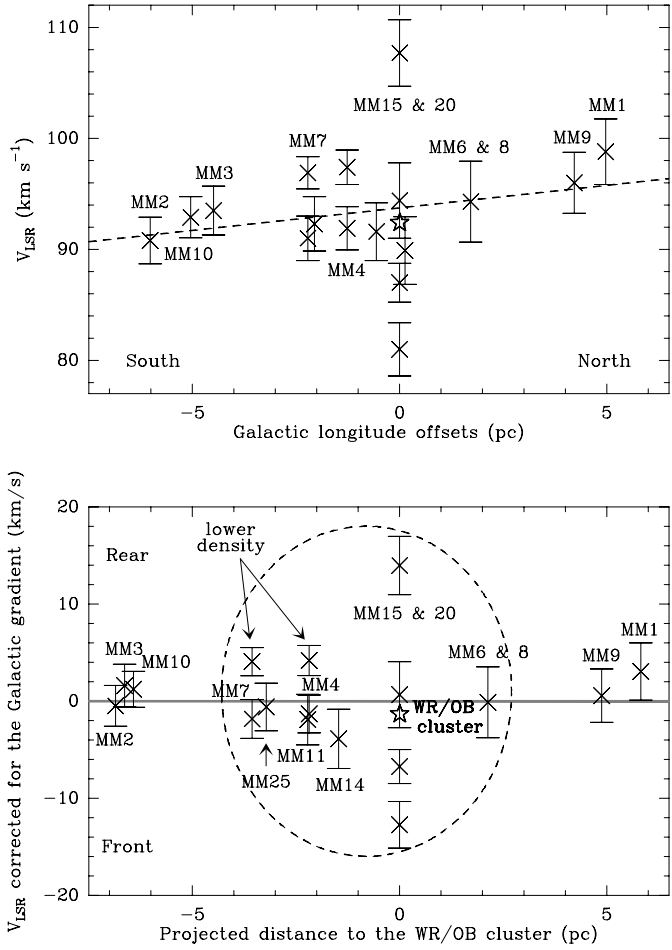}
\vskip -0.2cm
\caption[]{Systemic velocity of selected submillimeter fragments as 
a function of the Galactic longitude offsets ({\bf a}) and projected
distance ({\bf b}) from the WR/OB cluster.  The dashed line in {\bf
(a)} marks the least squares fit gradient which is subtracted from the
$\vlsr$ plotted in {\bf (b)}.  The location of the WR/OB cluster is
indicated with a star marker.  Error bars correspond to the line width
of the velocity components.  Note that in {\bf (b)}, the dispersion to
the zero line (gray thick line) and the line widths are larger within
the dashed ellipse.}
\label{f:vel_dist}
\end{figure}


\begin{thebibliography}{}
\bibitem[Andr\'e et al.(2000)Andr\'e, Ward-Thompson, \&
 Barsony]{awb00} Andr\'e, P., Ward-Thompson, D., Barsony, M. 2000, in
 Protostars \& Planets IV, ed. V. Mannings, A. Boss, \& S. Russell
 (Tucson: Univ. Arizona Press), 59
\bibitem[Balser et al.(2001)Balser, Goss, \& De Pree]{bal01} Balser,
 D. S., Goss, W. M., De Pree, C. G. 2001, \aj, 121, 371
\bibitem[Becker et al.(1994)]{beck94} Becker, R. H., White, R. L.,
 Helfand, D. J., Zoonematkermani, S.  1994, \apjs, 91, 347
\bibitem[Bieging et al.(1982)Bieging, Wilson, \& Downes]{bieg82}
 Bieging, J. H., Wilson, T. L., Downes, D. 1982, \aap S, 49, 607
\bibitem[Bertoldi(1989)]{bert89} Bertoldi, F. 1989, \apj, 346, 735
\bibitem[Bertoldi \& McKee(1992)]{bert92} Bertoldi, F., McKee,
 C. F. 1992, \apj, 395, 140
\bibitem[Beuther et al.(2002)]{beut02} Beuther, H., Schilke, P.,
 Menten, K. M., Motte, F., Sridharan, T. K., Wyrowski, F. 2002, \apj,
 566, 945
\bibitem[Blum et al.(1999)Blum, Damineli, \& Conti]{blum99} Blum,
 R. D., Damineli, A., Conti, P. S. 1999, \apj, 117, 1392
\bibitem[Bontemps et al.(2001)]{bonte01} Bontemps, S., Andr\'e, P.,
 Kaas, A. A., et al. 2001, \aap, 372, 173
\bibitem[Brand \& Blitz(1993)]{bran93} Brand, J., Blitz, L.  1993,
 \aap, 275, 67
\bibitem[Brand et al.(2001)]{bran01} Brand, J., Cesaroni, R., Palla, F., 
 Molinari, S. 2001, \aap, 370, 230
\bibitem[Brandl et al.(1999)]{bran99} Brandl, B., Brandner, W.,
 Eisenhauer, F., Moffat, A. F. J., Palla, F., Zinnecker, H.  1999,
 \aap, 352, L69
\bibitem[Braz \& Epchtein(1983)]{braz83} Braz, M. A., Epchtein,
 N. 1983, \aap S, 54, 167
\bibitem[Brogui\`ere et al.(1995)Brogui\`ere, Neri, \& Sievers]{brog95}
 Brogui\`ere, D., Neri, R., Sievers, A. 1995, NIC bolometer users guide
 (IRAM internal report)
\bibitem[Caswell et al.(1995)]{casw95} Caswell, J. L., Vaile, R. A.,
 Ellingsen, S. P., Whiteoak, J. B., Norris, R. P. 1995, \mnras, 272,
 96
\bibitem[Cesarsky et al.(1996)]{cesa96} Cesarsky, D., Lequeux, J.,
 Abergel, A., Perault, M., Palazzi, E., Madden, S., Tran, D. 1996,
 \aap, 315, 309
\bibitem[Churchwell(1999)]{chur99} Churchwell, E. 1999, in The Origin
 of Stars and Planetary Systems, ed. C. J. Lada \& N. D. Kylafis
 (Kluwer Academic Publishers), 515
\bibitem[D\'esert et al.(1990)D\'esert, Boulanger, \& Puget]{dese90}
 D\'esert, F.-X., Boulanger, F., Puget, J.L. 1990, \aap, 237, 215
\bibitem[Egan et al.(2001)]{egan01} Egan, M. P., Price, S. D., Moshir,
 M. M., Cohen, M., Tedesco, E., Murdock, T. L., Zweil, A., Burdick,
 S., Bonito, N., Gugliotti, G. M., Duszlak, J. 2001, VizieR On-line
 Data Catalog, originally published in: Air Force Research Lab.
 Technical Rep.
\bibitem[Emerson et al.(1979)Emerson, Klein, \& Haslam]{EKH} Emerson,
 D. T., Klein, U., Haslam, C. G. T. 1979, \aap, 76, 92
\bibitem[Figer et al.(1999)]{fige99} Figer, D. F., Kim, S. S., Morris,
 M., Serabyn, E., Rich, R. M., McLean, I. S. 1999, \apj, 525, 750
\bibitem[Garwood et al.(1988)]{garw88} Garwood, R. W., Perley, R. A.,
 Dickey, J. M., Murray, M. A. 1988, \aj, 96, 1655
\bibitem[Henning et al.(1995)Henning, Michel, \& Stognienko]{henn95}
 Henning, T., Michel, B., \& Stognienko, R. 1995, Planet. Space Sci.
 (Special issue: Dust, molecules and backgrounds), 43, 1333
\bibitem[Hobson et al.(1993)]{hobs93} Hobson, M. P., Padman, R.,
 Scott, P. F., Prestage, R. M., Ward-Thompson, D. 1993, \mnras, 264,
 1025
\bibitem[Hunter et al.(1996)Hunter, Benford, \& Serabyn]{hunt96}
 Hunter, T. R., Benford, D. J., Serabyn, E. 1996, \pasp, 108, 1042
\bibitem[Hunter et al.(2000)]{hunt00} Hunter, T. R., Churchwell, E.,
 Watson, C., Cox, P., Benford, D. J., Roelfsema, P. R. 2000, \aj, 119,
 2711
\bibitem[Jijina et al.(1999)Jijina, Myers, \& Adams]{jiji99} Jijina,
 J., Myers, P. C., Adams, F. C. 1999, \apjs, 125, 161
\bibitem[Kramer et al.(1998)]{kram98} Kramer, C., Stutzki, J., Rohrig,
 R., Corneliussen, U. 1998, \aap, 329, 249
\bibitem[Kreysa et al.(1998)]{krey98} Kreysa, E., Gem\"und, H. P.,
 Gromke, J., Haslam, C. G., Reichertz, L., Haller, E. E., Beeman,
 J. W., Hansen, V., Sievers, A., Zylka, R.  1998, in SPIE 3357,
 Advanced Technology MMW, Radio, and Terahertz Telescopes,
 ed. T. G. Phillips, 319
\bibitem[Lada(1987)]{lada87} Lada, C. J. 1987, in IAU Symp. 115, Star
 forming regions, ed. M. Peimbert \& J.  Jugaku, 1
\bibitem[Lefloch \& Lazareff(1994)]{lefl94} Lefloch, B., Lazareff, B.
 1994, \aap, 289, 559
\bibitem[Lester et al.(1985)]{lest85} Lester, D. F., Dinerstein, H. L., 
 Werner, M. W., Harvey, P. M., Evans, N. J., II, Brown, R. L. 1985, \apj, 
 296, 565
\bibitem[Liszt(1995)]{lisz95} Liszt, H. S. 1995, \aj, 109, 1204
\bibitem[Liszt et al.(1993)Liszt, Braun, \& Greisen]{lisz93} Liszt,
 H. S., Braun, R., Greisen, E. W. 1993, \aj, 106, 2349
\bibitem[Mathis(1990)]{math90}  Mathis, J. S. 1990, \araa, 28, 37
\bibitem[Mooney et al.(1995)]{moon95} Mooney, T., Sievers, A., Mezger,
 P. G., Solomon, P. M., Kreysa, E., Haslam, C. G. T., Lemke, R. 1995,
 \aap, 299, 869
\bibitem[Motte \& Andr\'e(2001a)]{ma01a} Motte, F., \& Andr\'e,
 P. 2001a, \aap, 365, 440
\bibitem[Motte \& Andr\'e(2001b)]{ma01} Motte, F., \& Andr\'e, P. 2001b,
 in ASP Conf. Ser. 243, From Darkness to Light, ed. T. Montmerle \&
 P. Andr\'e (San Francisco: ASP), 301
\bibitem[Motte et al.(1998)Motte, Andr\'e, \& Neri]{mott98} Motte, F.,
 Andr\'e, P., Neri, R. 1998, \aap, 336, 150
\bibitem[Motte et al.(2001)]{mott01} Motte, F., Andr\'e, P.,
 Ward-Thompson, D., Bontemps, S. 2001, \aap, 372, L41
\bibitem[Myers(1998)]{myer98} Myers, P. C. 1998, \apj, 496, L109
\bibitem[Nielbock et al.(2001)]{niel01} Nielbock, M., Chini, R.,
 J\"utte, M., Manthey, E. 2001, \aap, 377, 273
\bibitem[Ossenkopf \& Henning (1994)]{osse94} Ossenkopf, V., Henning,
 T. 1994, \aap, 291, 943
\bibitem[Plume et al.(1999)]{plum97} Plume, R., Jaffe, D. T., Evans,
 N. J., II, Mart\'{\i}n-Pintado, J., G\'omez-Gonz\'alez, J.  1997,
 \apj, 476, 730
\bibitem[Pound \& Blitz(1993)]{pb93} Pound, M. W., Blitz, L. 1993,
 \apj, 418, 328
\bibitem[Rainey et al.(1987)]{rain87} Rainey, R., White, G. J.,
 Gatley, I., Hayashi, S. S., Kaifu, N., Griffin, M. J., Monteiro,
 T. S., Cronin, T. S., Scivetti, A.  1987, \aap, 171, 252
\bibitem[Salpeter(1955)]{salp55} Salpeter, E. E. 1955, \apj, 121, 161
 \bibitem[Silk(1997)]{silk97} Silk, J. 1997, in AIP Conf. Ser. 393,
 Star Formation Near and Far VII, ed. S. S. Holt \& L. G. Mundy
 (Woodbury N. Y.: AIP), 3
\bibitem[Smith et al.(1978)Smith, Biermann, \& Mezger]{smit78} Smith,
 L. F., Biermann, P., Mezger, P. G. 1978, \aap, 66, 65
\bibitem[Sridharan et al.(2002)]{srid02} Sridharan, T. K., Beuther, H., 
 Schilke, P., Menten, K. M., Wyrowski, F. 2002, \apj, 566, 931
\bibitem[Starck et al.(1998)Starck, Murtagh, \& Bijaoui]{star98}
 Starck, J.-L., Murtagh, F., Bijaoui, A. 1998, Image Processing and
 Data Analysis: The Multiscale Approach, Cambridge: Cambridge
 Univ. Press
\bibitem[Tapia et al. (2001)]{tapi01} Tapia, M., Bohigas, J., P\'erez,
 B., Roth, M., Ruiz, M. T. 2001, RMxAA, 37, 39
\bibitem[Valdettaro et al.(2001)]{vald01} Valdettaro, R., Palla, F.,
 Brand, J., Cesaroni, R., Comoretto, G., Di Franco, S., Felli, M.,
 Natale, E., Palagi, F., Panella, D., Tofani, G. 2001, \aap, 368, 845
\bibitem[van der Tak(2002)]{vdt02} van der Tak, F. F. S. 2002, in ASP
 Conf. Ser. 267, Hot Star Workshop III: The Earliest Phases of Massive
 Star Birth, ed. P. A.  Crowther (San Francisco: ASP), 33
\bibitem[Walsh et al.(1998)]{wals98} Walsh, A. J., Burton, M. G.,
 Hyland, A. R., Robinson, G.  1998, \mnras, 301, 640
\bibitem[Wilking \& Lada(1983)]{wilk83} Wilking, B. A., Lada, C. J
 1983, \apj, 274, 698
\bibitem[Wilson et al.(1970)]{wils70} Wilson, T. L., Mezger, P. G., 
 Gardner, F. F., Milne, D. K. 1970, \aap, 6, 364
\bibitem[Wood \& Churchwell(1989)]{wood89} Wood, D. O. S., Churchwell,
 E.  1989, \apj, 340, 265
\bibitem[Zoonematkermani et al.(1990)]{zoot90} Zoonematkermani, S.,
 Helfand, D. J., Becker, R. H., White, R. L., Perley, R. A. 1990,
 \apjs, 74, 181
\end{thebibliography}
\end{document}